\documentclass[10.5pt]{report}
\usepackage[utf8]{inputenc}
\usepackage[pdftex]{graphicx}
\usepackage{multicol}
\usepackage{pdfpages}
\usepackage{authblk}

\graphicspath{[figures/}

\usepackage{etoolbox}
\makeatletter
\patchcmd{\chapter}{\if@openright\cleardoublepage\else\clearpage\fi}{}{}{}
\makeatother

\title{
{TrustSECO: An Interview Survey\\ into Software Trust}\\\
{\large Universiteit Utrecht}\\\
{\includegraphics{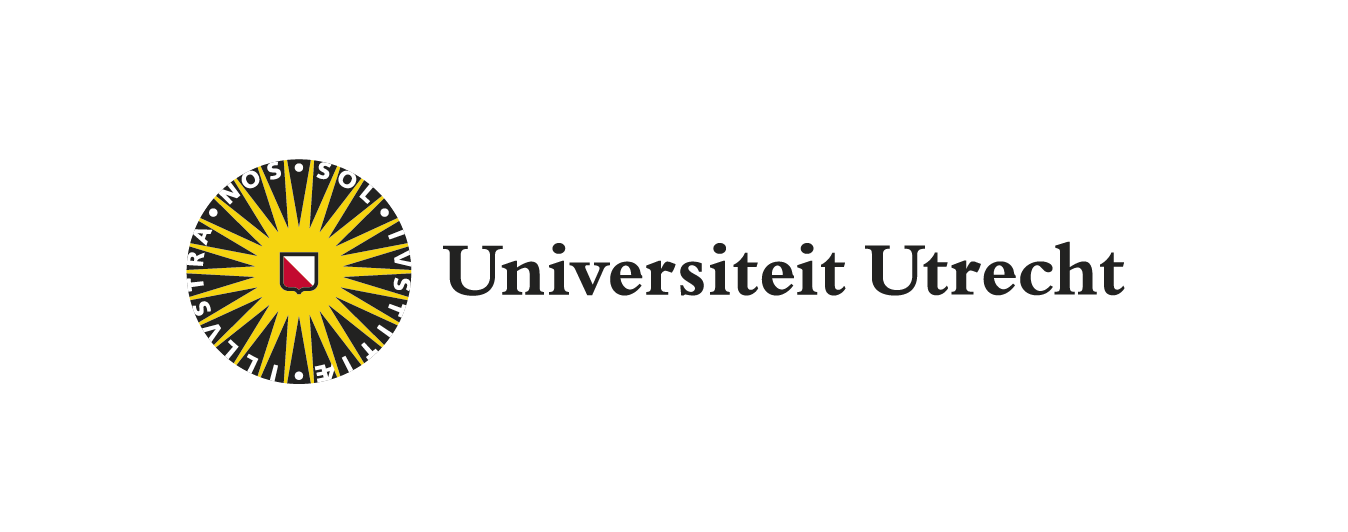}}

}
\author{Floris Jansen  -  6002919}

\affil{Dr. R.L. Jansen}
\affil{Dr. G.C. van de Weerd}
\date{January 2021}

\affil{Department of Information and Computing Sciences}

\begin{document}

\maketitle

\chapter*{Abstract}
The software ecosystem is a trust-rich part of the world. Collaboratively, software engineers trust major hubs in the ecosystem, such as package managers, repository services, and programming language ecosystems. This trust, however, is often broken by vulnerabilities, ransomware, and abuse from malignant actors. 

But what is trust? In this paper we explore, through twelve in-depth interviews with software engineers, how they perceive trust in their daily work. From the interviews we conclude three things. First, software engineers make a distinction between an adoption factor and a trust factor when selecting a package. Secondly, while in literature mostly technical factors are considered as the main trust factors, the software engineers in this study conclude that organizational factors are more important. Finally, we find that different kinds of software engineers require different views on trust, and that it is impossible to create one unified perception of trust.
\\ \\
Keywords:  software ecosystem trust, empirical software engineering, TrustSECO, external software adoption, cross-sectional exploratory interview analysis,  trust perception.

\newpage

\tableofcontents

\newpage
\chapter{Introduction}
Software engineers use software packages for creating new solutions from different package managers. \cite{mojica2014scale} \cite{nguyen2020crossrec}. There is a significant amount of implicit trust in these packages. While the packages could easily be compromised, software engineers assume that as the package comes from a reliable source, it must be trustworthy. This is not always the case \cite{duan2020measuring}. Moreover, there are several attack vectors that can  compromise a package like registry exploitation of typo squatting \cite{trust2020seco}. 

Before the factors that constitute trust in software packages and package repositories are looked upon, one must look at how software engineers choose software packages and how trust is gained. Moreover, what the impact factors are that influence this. The TrustSECO project aims to uncover all the factors that influence the trust that software engineers have in software packages. In order to uncover these impact factors, a survey will be developed that has as its main aim to uncover how software engineers perceive trust. \cite{vargas2020selecting} has done similar research and found 26 factors that influence the selection process of software packages. Rather than looking at the whole selection process, this research will just focus on the trust aspect of that selection process.
This will be done by analyzing existing literature and the results of cross-sectional interviews with experts. This information will then be the basis for a large scale survey. This research thus sets out to find what trust factors influence the decision to choose software packages. 
\\
\\

\newpage
\chapter{Framework}
\section{Trust}
One of the most important aspects of collaboration is trust \cite{bunduchi2013trust}. Moreover, trust creates the basis in decision making for the usage of long term product use. \cite{cho2015trust}. Using external software is just that, collaboration. In order to find out what factors induce trust in packages. One must take a look at the term trust. The term is widely used in computer science and has many different definitions across the spectrum. \cite{artz2007survey}. 

There has been a lot of research on this subject that lead to the following three most general and common definitions of trust \cite{artz2007survey}.
\begin{itemize}
    \item “Trust is a subjective expectation an agent has about another's future behavior based on the history of their encounters.” \cite{mui2002computational}
    \item “Trust is the firm belief in the competence of an entity to act dependably, securely, and reliably within a specified context.”~\cite{grandison2000survey}
    \item “Trust of a party A to a party B for a service X is the measurable belief of A in that B behaves dependably for a specified period within a specified context (in relation to service X).”~\cite{olmedilla2006security}
\end{itemize}

The first definition is a reputation based one because it concerns the producer of the software rather than the software itself. While the producer is highly relevant in gaining trust in a software package. The producer of the package will not be more important than the product or service itself. This is because this research looks at what factors induce trust in software packages, not at the factors that induce trust in the entities that develop the software. 

The second definition is a definition that suits this research better since it concerns the belief in the competence of the software. Therefore the characteristics for the software have to ensure that it acts dependably, securely and reliably. These characteristics will induce a list of factors that ensure that the system acts this way.

The third and final definition of trust concerns a service from a party to another party. This trust also comes from the belief in the product or service, rather than the party that produces this. Therefore this definition of trust will also suit this research.

The definition of trust that will be leading for this research is:\\ \textit{``Trust of a party A to a party B for a service X is the measurable belief of A in that B behaves dependably for a specified period within a specified context (in relation to service X).''}\\ For this definition specifies the collaboration of two parties regarding a specific service or product. Thus the trust influence of the developer is not discarded while the focus is on the product, as is the scope for this research. 

\section{Bias}
Bias is a form of error that can affect research. \cite{sica2006bias}. For this research, this may occur in the form of selection bias. Selection bias may occur on the participants in the interview if they do not represent the general ideas and thoughts of the average participant. \cite{hernan2004structural}  Therefore it is important that the participants have different perspectives and roles on the same matter. This thus leads to participants in different organizations across different fields across different jobs and functions.

Another more intuitive form of bias present will be when several trust factors will be discussed in the interviews, personal bias. After all each participant has their own personal experiences with packages. Minimizing this bias will be next to impossible, since the participant might not even be aware of this bias. However this personal bias will be decreased by first letting the interviewee answer some more technical impersonal questions before diving into the personal part.  
\\
\\

\newpage
\chapter{Research Methods}
\begin{figure}[htbp]
\hspace*{-2cm}
\includegraphics[scale=0.22]{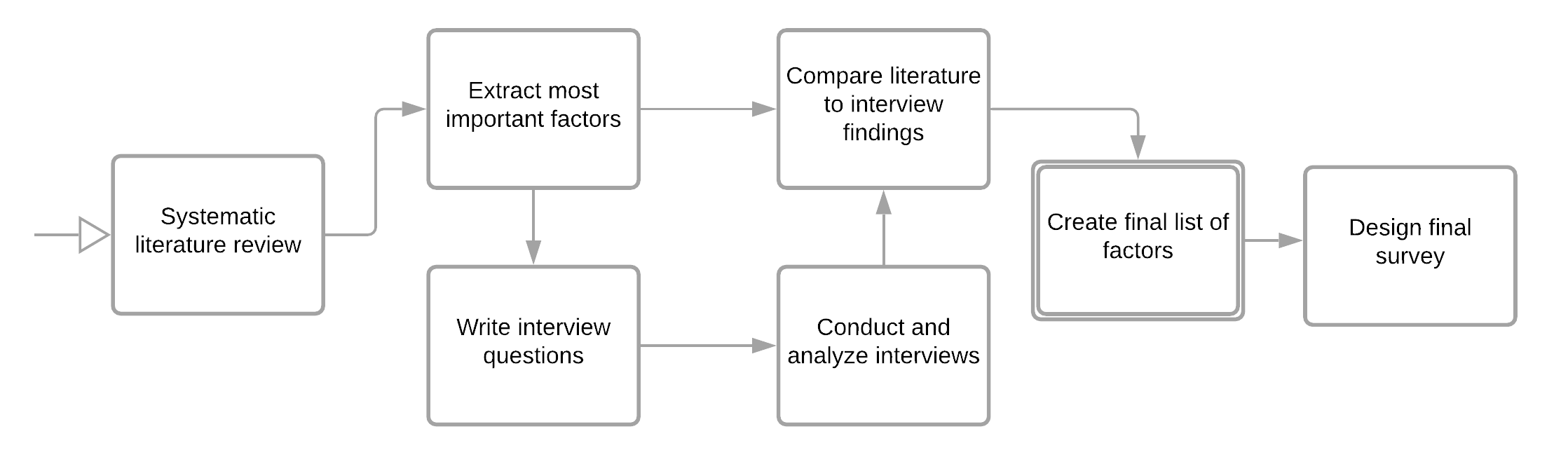}
\caption{Research process}
    \label{researchprocess}
\end{figure}

To find what factors influence trust in software packages, one must first find all the adoption factors that are considered when making such decision. After this, a selection can be made of the factors that actually contribute to trusting a package.
In order to explore the current state of impact factors already present in literature, a literature study will be conducted. These results will be compared with the results from the cross-sectional interviews with software engineers to complement this current literature.
This process is shown in figure \ref{researchprocess}.

\newpage
\section{Literature study}
The literature study will be based on the SLR done by the TrustSECO te am. This SLR will to try uncover the already existing factors in literature.  Various search queries will be entered in the following search engines:

\begin{itemize}
   \item Google scholar   
   \item IEEE Xplore
   \item ScienceDirect
   \item Jstor   
\end{itemize}
Figure 3.2 illustrates all the used search queries. By systematically using all the combinations as search queries, all the relevant literature is found. This relevant literature will be stored and analyzed for trust factors. 
\begin{figure}[ht!]
\centering
    \includegraphics[scale=0.7]{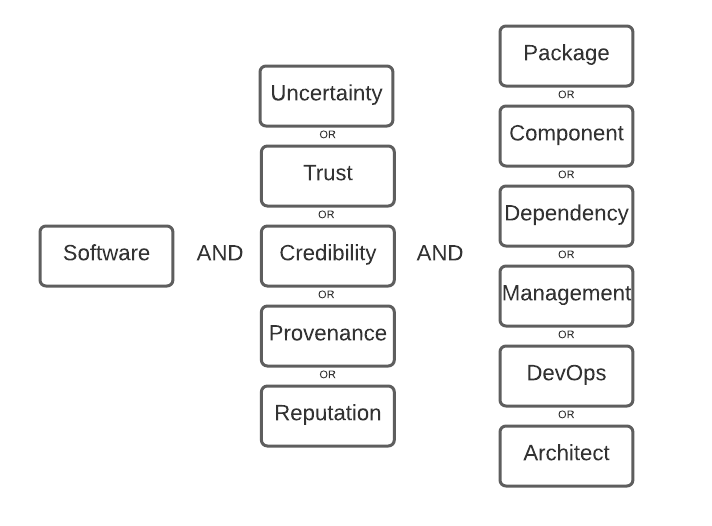}
    \\
    \caption{Search terms}

\end{figure}

The combination of these search terms will result in a list of articles. This list will be narrowed down through exclusion and inclusion rules. These are as follows:

\begin{itemize}
  \item Literature should be about open source software 
  \item Literature should list at least one impact factors on adopting open source software or one factor for gaining trust in a package
  \item Literature should be public and accessible
\end{itemize}

The last round of elimination will be done through abstract analysis of the articles. This will result in a final list of articles to be analyzed.These articles can be found in appendix \ref{SLR} These articles will be scoured for adoption and trust factors and will be categorised as follows:

\begin{itemize}
  \item Technical factors
  \item Organizational factors
  \item Economic factors
\end{itemize}

 The categorization is based on \cite{vargas2020selecting} which holds an explanation as to why a certain factor is categorized as such. Technical factors are factors related to the release process, code quality attributes and the functionality. Organizational factors concern the individual perception, community around the project and other aspects of the organization where the package is developed. The economical aspects cover the financial aspect of package selection like licences, total cost of ownership and risks.

\section{Interviews}
The literature review will provide a solid foundation of knowledge to start creating interview questions. Interviews can be used to get detailed personal experiences and thought processes in the selection process~\cite{hiller2004interviewee}. These semi-structured interviews shall be conducted with software engineers, DevOps, architects and other experts (see table \ref{participantoverview}) The choice for semi-structured is made since this gives the freedom for the interviewee to really elaborate on personal experiences. In addition, this allows the interviewer to create a new line of questioning based on those experiences~\cite{hove2005experiences}. These interviews are held in English or Dutch. Since the literature provided English factors, a list of translated terms will be provided to ensure that the interviews in Dutch will not yield different results because of difference in understanding in certain terminology across these languages.
Prior to the interview, an interview protocol is created based on \cite{jacob2012writing}. 

These semi-structured interviews are a form of exploratory research. The subject of exploratory research has been named by among others \cite{glaser2017discovery} with the discovery of grounded theory.  The goal of exploratory research has been to form hypotheses rather than the testing of hypotheses \cite{kothari2004research}. This is the case for this research since the objective is to define a hypothesis describing which factors lead to trusting a software package.

Each interview discusses roughly 9 questions regarding software trust. In order to ensure the participants have the required knowledge to answer the interview questions, certain standards have to be met, namely:
\begin{itemize}
  \item The participant needs to speak English or Dutch fluently
  \item The participant has to have been involved with the selection process of software packages
  \item The participant needs at least three years of relevant working experience
\end{itemize}

\begin{table}[h!]
\footnotesize
\resizebox{1\textwidth}{!}{\begin{minipage}{\textwidth}
\begin{tabular}{|l|l|l|r|l|r|}
\hline
\textbf{Nr} &
  \textbf{Organisation} &
  \textbf{Sector} &
  \textbf{Size} &
  \textbf{Function/role} &
  \multicolumn{1}{l|}{\textbf{Experience}} \\ \hline
P1  & Triodos Bank               & Banking              & 1000+ & Software Engineer       & 9  \\ \hline
{\color[HTML]{000000} P2} &
  {\color[HTML]{000000} Bol.com} &
  E-commerce &
  1500+ &
  Product Owner &
  6 \\ \hline
P3  & Universiteit Twente        & Education            & 3000+ & Tech Product owner & 20 \\ \hline
P4  & Keylane                    & Insurance    & 400+  & Software Engineer       & 7  \\ \hline
P5  & Xebia                      & IT Consultancy       & 300+  & DevSecOps               & 26 \\ \hline
P6  & BOC Group            & IT Consultancy       & 200+  & Web developer           & 5  \\ \hline
P7  & Channable                  & Marketing            & 100+  & DevOps                  & 5  \\ \hline
P8  & Ministry of Defence & Military             & 3000+ & Software Engineer       & 12 \\ \hline
P9  & NOS                        & News                 & 600+  & Software Engineer       & 7  \\ \hline
P10 & Gemboxx                    & Software dev               & 10+   & Software Engineer       & 5  \\ \hline
P11 & Sogeti                     & Software dev & 2500+ & Software Architect      & 23 \\ \hline
P12 & Grasple & Ed-sec & 10+     & Software Engineer                                & 16 \\ \hline
\end{tabular}

\caption{Participant overview}
\label{participantoverview}
\end{minipage}}

\end{table}

\newpage
The responses give insight as to why certain packages are chosen and the factors that contributed to this decision. Moreover to answer: 
\\  \\
\textbf{How do software engineers develop trust in a package?} 
\\ \\
In order to answer this question, the interviews must first provide answer to the following questions:
\begin{itemize}
    \item What factors are important when selecting external software packages, what is the protocol?
    \item Which of the selecting factors contribute to gaining trust in a package?
    \item How do personal aspects influence trust in packages? 
\end{itemize}

The first and second sub question may seem alike. However the literature review has shown that the factors that induce trust are a subset of the factors that influence the choice of a package. 
These sub questions will each be answered through a section in the interview. Each section contains the actual interview questions that the interviewee will be answering. Section one will contain Q1...Q4, section two will contain Q5 and Q6 and section three will contain Q7...Q9. The full interview structure can be seen in appendix \ref{interviewlayout}.

These interviews are recorded and transcribed. After transcript approval from the participant a proper analysis will conclude the most important trust factors. This will create a list of factors that have quotes to back them up.\\

\chapter{Results}
\section{Literature study}
The literature study will first take a look at what factors influence the adoption of certain software packages, these will be categorized based on the research of \cite{vargas2020selecting}. What follows is an analysis of which of those factors actually lead to more trust in a software package. 

\cite{sanches2018open} was found during the SLR. This research describes a massive systematic literature review that was conducted to find the most important adoption factors when choosing open source software packages. This research will be setting the stage for the first part of the literature review. The factors will be discussed briefly to provide context and eventually be analyzed to see if they also induce trust.

\subsection{Technical adoption factors}
The technical factors are encountered in literature very frequently. These factors are concerning the technical aspects of software packages and are found in 49 of 54 pieces of literature. Impact factors such as compatibility with software, reliability usability and customization are the most common and thus have the highest importance. A brief description of the impact factors found according to \cite{wheeler2011evaluate} to get a good understanding how important certain factors can be:
\begin{itemize}
    \item \textbf{Compatibility:} This impact factor refers to the degree to which a piece of software integrates with existing software. Also whether or not additional programs are required to adopt this piece of software.
    \item \textbf{Reliability:} This factor measures if the software gives the wanted answers. This could be compared to availability. It is not a quantitative property and thus hard to measure. 
    \item \textbf{Usability:} This describes how intuitive the program is for the user. This impacts the difficulty of the software to learn. When software is very usable it will be adopted faster. 
    \item \textbf{Customization:} Customization is the degree to which a component can be changed to do something it could not do before. Configure changes to its initial configuration.
    \item \textbf{Documentation:} This refers to the available qualitative and quantitative documentation on a software package, this impact factor also falls under the 'support' impact factor, however it is heavily discussed in literature since it describes the technical capabilities of a package. 
    \item \textbf{Re-usability:} Re-usability concerns the quantity of actual code that can be reused in different fashions than the particular library uses it for. The more general a piece of software is, the more goals it can serve.
    \item \textbf{Triability:} This factor describes the ease of implementation in a system. When this is the case, several other factors can be tested quite easily and thus benefit the package.
    \item \textbf{Portability:} This is the least mentioned factor and thus carries the least importance of the technical factors.
    It concerns the ease to deploy the package in multiple different systems.
\end{itemize}

\begin{table}[h!]
\begin{tabular}{|l|c|}
\hline
\textbf{Technical adoption factors} &  \textbf{Mentioned number of times} \\ \hline
\textit{Compatibility}              & 34                                 \\ \hline
\textit{Reliability}                & 23                                 \\ \hline
\textit{Usability}                  & 17                                 \\ \hline
\textit{Customization}              & 17                                 \\ \hline
\textit{Documentation}              & 12                                 \\ \hline
\textit{Maintainability}            & 12                                 \\ \hline
\textit{Re-usability}               & 8                                  \\ \hline
\textit{Triability}                 & 9                                  \\ \hline
\textit{Portability}                & 6                                  \\ \hline
\end{tabular}
\caption{The 9 technical factors found in literature according to \cite{sanches2018open}}
\label{techadoptionfactors2}
\end{table}

\subsection{Human/organizational adoption factors}
This second category of factors concerns the organizational aspects of the entity. This category of factors did also have a dominant presence in literature. Overall this was a bit less than the technical factors however, this category does contain the most important impact factor: "support". This was named 45 times out of the 54 pieces. 

\begin{itemize}
    \item \textbf{Support:} This impact factor also contains 'technicalities' like documentation and release frequency. It is a very broad impact factor since it contains a variety of factors that express the need to have a backup plan when the organization does not have the technical skills to solve certain problems. This ranges from documentation to the community.  
    \item \textbf{Training:} A presence of sufficient training material ensures that technical staff can learn to fix problems themselves. It also ensures the users know the package is easy to learn and does not have to figure out the technical details themselves.
    \item \textbf{Top management support:} Some organizations' strategy can also impact the decision in which packages to adopt. The users select packages based on policies that the company has, e.g. privacy policy or information protection. \cite{vargas2020selecting}
    \item \textbf{Attitude towards change:} This factor describes how employees are looking towards the adoption of new packages. 
    \item \textbf{Case studies of FLOSS adoption:} This factor describes the success of implementing new software packages. When a package is widely adopted  it gains in reputation. Hence it can influence the decision process.
    \item \textbf{Time adoption:} This factor concerns the total time it takes to implement a package. The longer it takes to fully adapt to new software the less appealing it is.  
    \item \textbf{Centrality IT:} This factor describes the dependency of the organization to the new software. Since the proper implementation of more important systems is way more urgent less important systems.
    \item \textbf{Business process engineering:}  This describes "when an organization is changing  its internal business processes due to any particular circumstance(e.g. quality improvement, organizational restructure)." This factor is only mentioned once throughout all literature and makes it the least named and thus least important.
\end{itemize}
\begin{table}[h!]
\begin{tabular}{|l|c|}
\hline
\textbf{Organizational adoption factors} &  \textbf{Mentioned number of times} \\ \hline
\textit{Support}              & 45                            \\ \hline
\textit{Training}                & 25                                \\ \hline
\textit{Vendor locks-in}              &     13                         \\ \hline
\textit{Top management support}                  & 10       
\\ \hline
\textit{Attitude}              & 6                                 \\ \hline
\textit{Centrality IT}            & 3                              \\ \hline
\textit{Time of adoption}               & 2                                \\ \hline
\textit{Case studies}                 & 2                                \\ \hline
\textit{Business process re-engineering}                & 1                                \\ \hline
\end{tabular}
\caption{The 9 Organizational factors found in literature according to \cite{sanches2018open}}
\label{orgadoptionfactors2}
\end{table}

\subsection{Economic adoption factors}
The last category of factors concerns the evaluation of economic factors. These factors are have the least dominance in literature however are still important for software practitioners to look upon. 

\begin{itemize}
    \item \textbf{Total cost of ownership:} Total cost of ownership contains all the costs related to the software package. This includes licensing, however also includes operational and support costs.
    \item \textbf{Licensing  cost:} Is a part of the total cost of ownership. This is the main cost that practitioners think about when discussing software costs, and concerns the cost of obtaining a particular license.
    \item \textbf{Operational cost:} This consists of three aspects: the cost to change from systems, the maintenance of the solution and the costs to implement the solution.
    \item \textbf{Support cost:} These are the costs related to external support as well as keeping the system updated. This is only referenced twice in literature and thus not a very influencing factor.
\end{itemize}

\begin{table}[h!]
\begin{tabular}{|l|c|}
\hline
\textbf{Economical adoption factors} &  \textbf{Mentioned number of times} \\ \hline
\textit{Total cost of ownership}              & 10                                \\ \hline
\textit{License cost}                & 16                               \\ \hline
\textit{Operational cost}                  & 4                               \\ \hline
\textit{Support cost}              & 2                                \\ \hline

\end{tabular}
\caption{The 4 economical factors found in literature according to \cite{sanches2018open}}
\label{economicaladoptionfactors}
\end{table}

\subsection{Trust factors}
The factors that influence the adoption of open source software may differ from the factors that induce trust. Factors like functionality and reliability are constantly ranked as high for building trust, next to some other technical and organizational factors. \cite{del2011survey}. However it seems that the trustworthiness of a system can supersede other adoption factors. \cite{bernstein2005trustworthy}.

The research done by \cite{del2011survey} provides a list of factors that are believed to affect trustworthiness the most according to it's interviewees. The two most important factors are "Reliability" and "Degree to which an OSS product satisfies/covers functional requirements the most". These factors are the two biggest technical impact factors in this research.

\begin{table}[h!]
\resizebox{0.92\textwidth}{!}{\begin{minipage}{\textwidth}
\hskip2.9cm\begin{tabular}{|l|c|}
\hline
\textbf{Technical trust factor} &  \textbf{Rank} \\ \hline
\textit{Reliability}              & 8                                 \\ \hline
\textit{Alignment with software}                & 8                               \\ \hline
\textit{Interoperability}                  & 7                                 \\ \hline
\textit{Maintainability}              & 6                                 \\ \hline
\textit{Standard compliance}              & 6                                 \\ \hline
\textit{Performance}            & 5                             \\ \hline
\textit{Usability}               & 5                                  \\ \hline
\textit{Security}                 & 5                                 \\ \hline
\textit{Portability}                & 4                                 \\ \hline
\textit{Reusability}              & 4                                   \\ \hline
\textit{Modularity}                & 4                                  \\ \hline
\textit{Standard architecture}                & 4                                 \\ \hline
\textit{Human interface}                & 3                                \\ \hline
\textit{Complexity}                & 2                                  \\ \hline
\textit{Patterns}                   & 2\\ \hline
\textit{Self-containedness}                & 2                                 \\ \hline
\textit{Size}                & 1                                  \\ \hline
\end{tabular}
\caption{The technical attributes ranked on inducing the most trust according to \cite{del2011survey}}
\end{minipage}}
\end{table}

The most important adoption factor for this research was support. However on the list of trustworthiness "Short-term support" is number fourteen on the list. This illustrates that even though for adoption it is the most important factor, for trusting certain software the technical factors are more important. 

Another remarkable difference between adoption and trustworthiness factors is that the economic factors are mentioned 12(licence) and 13(total cost of ownership) times out of the 31 articles. Yet in the list of trustworthiness factors there is only one economical factor present out of the 37 named trust factors. That is also why there is no table for economical trust factors.

On the contrary, a factor that is in the second highest scoring group for trustworthiness is customer "Satisfaction". While it is trivial that this factor is important for most businesses, it is not explicitly named in literature about adopting open source software. This could be since it is often used to create a solution for customers rather than being the solution itself. 

\begin{table}[h!]
\resizebox{1.0\textwidth}{!}{\begin{minipage}{\textwidth}
\hskip2.3cm
\begin{tabular}{|l|c|}

\hline

\textbf{Organizational trust factor} &  \textbf{Rank} \\ \hline
\textit{Documentation}              & 7                                \\ \hline
\textit{Mid-/long existent community}                & 6                                \\ \hline
\textit{Community experience}                  & 5                                \\ \hline
\textit{Short-term support}              & 5                                 \\ \hline
\textit{Availability of support tools}              & 4                                 \\ \hline
\textit{Environmental issues}            & 4                                 \\ \hline
\textit{Availability of best practices}               & 3                                  \\ \hline
\textit{Programming language uniformity}                 & 3                                  \\ \hline
\textit{Training}                & 2                                  \\ \hline
\textit{Benchmarks/ test suites}                & 2                                  \\ \hline
\textit{Organization}                & 2                                  \\ \hline
\textit{Reputation}                & 2                                  \\ \hline
\textit{Distribution channel}                & 1                                  \\ \hline
\end{tabular}
\caption{The organizational attributes ranked on inducing the most trust according to \cite{del2011survey}}
\end{minipage}}
\end{table}

\newpage
\section{Interviews}
The interview protocol is divided into 3 sections. These sections will be discussed separately and will answer the sub question set for that section.

\subsection{Section 1, adoption factors and selection procedure}
Section 1 of the interview consists of two separate parts containing four questions. The answers to these questions aim to answer the following sub question: What factors are important when selecting external software packages, what is the protocol? 
\subsubsection{Selection procedure}
In order to get all the relevant information from the interviewees. The focus of the first section of the interview was to find out the procedure when selecting other packages, as well as finding out all the relevant factors. This procedure is highly relevant because it describes what role a developer has when selecting packages and the processes they have to go through. This sets the stage on their perspective on this matter.  
When asked about the procedure, all participants somehow mentioned that this was dependant of the situation. Important and sensitive project require delicate measures and a more sophisticated procedure. Where less important projects would not require these. After that, the participants were asked to describe the typical procedure when selecting packages in their company. In some occasions there was a formal protocol when selecting packages (P1, P8, P12). These procedures stated the responsibilities of certain actors within the company and delegated roles. These procedures were set because of the impact a vulnerability could lead to. P8: \textit{"So if we are want to use a package it first arrives in the 'demilitarized-zone' , then we conduct our analysis. Both by hand and with automated systems, and they tell us whether or not it meets all the set requirements."}

A more common practice amongst the interviewees is an informal procedure. (P2, P3, P5..P7, P9..P11). This was not a clear set of rules and responsibilities. However there is a common practice when it comes to selecting packages. The most important aspect across all informal procedures is that a decision to adopt a package is always discussed within the development team. P3: \textit{"We do this by reviewing each others work, and review of the library's code. Each code change is monitored by the team."} In two occasions this sometimes has to be explicitly discussed with an architect or DevOps (P7, P11). This social security ensures that all code is reviewed and no code is accepted without being looked by at least two members of the team. 

In addition to discussing these matters within the team, there is one other aspect that is heavily relied on when selecting packages with an informal selection process. This is the common sense and self-awareness of the developer. P:6 \textit{"... rules are not written in stone to be honest, and we just do what we think is right."} While this is trivial, it is named very specifically by P3, P5, P6, P10. The participants that had a formal process did not mention this, since the risk a development team could have is safeguarded by the formal procedure. So for the informal procedure this implies that there is a lot of implicit trust amongst team members, this can bring a significant risk to the project. 

When looking at the participants who either had a formal or informal protocol, one can see that P4 has none. This is because this participant has not operated in a development team. However when continuing the line of questioning, after a while he revealed he had some sort of self-made protocol in his mind for selecting packages. Through time a mind model was created on how to approach these matters.

\subsubsection{Adoption factors}
All the adoption factors that were found during the interviews are divided into the following categories:
\begin{itemize}
    \item {Technical adoption factors} 
    \item {Organizational adoption factors}
    \item {Economic adoption factors}
\end{itemize}

\begin{table}[h!]
\centering
\footnotesize
\resizebox{0.92\textwidth}{!}{\begin{minipage}{\textwidth}
\hskip-0.6cm\begin{tabular}{lcccccccccccc}
\hline
 & P1 & P2 & P3 & P4 & P5 & P6 & P7 & P8 & P9 & P10 & P11 & P12 \\ \hline
\textbf{Technical} &  &  &  &  &  &  &  &  &  &  &  &  \\ \hline
\textit{Compatibility} &  & • &  &  &  & • & • &  & • & • &  & • \\
\textit{Documentation} &  & • & • & • & • & • & • & • &  & • &  & • \\
\textit{Complexity} & • &  & • &  & • &  &  & • &  &  &  &  \\
\textit{Security} &  &  &  &  &  &  & • & • &  & • &  &  \\
\textit{Source code} &  & • &  &  &  &  &  &  & • &  &  &  \\
\textit{Code quality} &  & • &  &  &  &  &  &  & • &  &  & • \\
\textit{Customization} &  &  & • &  &  &  &  &  &  &  &  &  \\
\textit{Usability} &  &  &  &  &  & • &  &  &  &  &  &  \\ \hline
\textbf{Organizational} & \textbf{} & \textbf{} & \textbf{} & \textbf{} & \textbf{} & \textbf{} & \textbf{} & \textbf{} & \textbf{} & \textbf{} & \textbf{} & \textbf{} \\ \hline
\textit{Active maintenance} & • &  & • & • & • & • &  &  & • &  & • & • \\
\textit{Number of contributors} & • &  &  &  &  &  &  &  & • &  & • &  \\
\textit{Contributor process} &  &  &  &  & • &  &  & • &  &  &  &  \\
\textit{Git Issues} & • &  &  & • & • & • &  &  & • &  &  & • \\
\textit{Number of users} & • & • & • &  &  & • &  &  &  & • &  &  \\
\textit{Backing company} &  & • &  & • &  &  &  &  &  &  & • &  \\
\textit{Ease of integration} & • & • & • &  &  & • &  &  &  & • &  & • \\
\textit{Community} &  &  &  &  & • & • & • &  &  &  &  &  \\
\textit{Known vulnerabilities} &  &  &  &  &  &  & • & • &  &  &  &  \\
\textit{Reputation} &  & • &  &  &  &  & • & • &  &  &  &  \\
\textit{Support} &  &  &  & • & • &  &  &  &  &  & • &  \\
\textit{Git stars} &  &  &  & • &  &  &  &  &  &  &  &  \\
\textit{Knowledge of team} &  & • &  &  &  &  &  &  &  &  &  &  \\ \hline
\textbf{Economical} & \textbf{} & \textbf{} & \textbf{} & \textbf{} & \textbf{} & \textbf{} & \textbf{} & \textbf{} & \textbf{} & \textbf{} & \textbf{} & \textbf{} \\ \hline
Licence &  &  &  &  &  &  &  & • &  &  & • & • \\
\textit{Total cost of ownership} & • &  &  &  &  &  &  &  &  & • &  &  \\ \hline
\end{tabular}
\caption{Adoption factors overview}
\end{minipage}}
\end{table}

One may notice that the table holds more concrete examples of factors rather than the factors itself. Another noticeable aspect is that some of the named aspects may overlap in their meaning. E.g. the 'Number of users' and 'Number of contributors' are two separate aspects, where 'Community' also holds the number of users and contributors. This is because the majority of the participants described the actual actions and thought processes of why they would look at a certain aspect. Therefore it would not be correct to categorize those specific aspects in a category, even though they do fall under that category.

A brief description of the specific aspects will now follow, along with a list of quotes to back them up.

\subsubsection{Technical factors}
Appendix \ref{techadoptionfactors2} holds the found technical aspects that influence the decision of which package to choose. After every aspect there is at least one quote from experts that support this. The following technical aspects were found in the interviews that influenced the interviewees as follows:
\begin{itemize}
   \item \textit{Compatibility}  - The package in question must align with the current framework or build. This is an aspect that has to hold before any other characteristics are considered.
   \item \textit{Documentation} - Dependent on the size and the complexity, documentation can play a big role in the decision of a package. This can also illustrates a picture on how serious the project is.
   \item \textit{Complexity} - A package must not be too complex for the problem that is trying to solve, and should make coding easier.
   \item \textit{Security} - A package should be secure, however there is no clear ways to determine this were mentioned. 
   \item \textit{Source code openness} - If the source code is available, a global scan can give an indication of the quality of the code and what is does. 
   \item \textit{Code quality} - The code should do what is claims it does and nothing more. Tests can help to validate this.
   \item \textit{Customization} - If a package does not fully cover your needed functionalities, it is very convenient is small adjustments can be made to get full coverage.
   \item \textit{Usability} - If a package is easy to use, it will contribute to the selection of that package.
\end{itemize}

\subsubsection{Organizational factors}
Appendix \ref{orgadoptionfactors2} holds the found Organizational aspects that influence the decision of which package to choose. After every aspect there is at least one quote from experts that support this. The following organizational aspects were found in the interviews that influenced the interviewees as follows:

\begin{itemize}
   \item \textit{Active maintenance}  - If a package is actively maintained, and has been for a period of time. It illustrates a stability, which is an important factor in the selection process.
   \item \textit{Number of contributors} -  When a lot of contributors work on a project it increases the stability of the package as well because if one of the contributors quits, there are enough others to take over the workload.
   \item \textit{Contributor process} - This describes the ease of becoming a contributor for a project. There is no general way of finding this out, however one can get a feel of how hard it is to make contributions to a project. If this is very easy, it also means that people with a bad agenda can do this and potentially create a vulnerability in an upcoming release.
   \item \textit{Git Issues} - Dependant on the type of Git issues, this can influence the choice of a package a lot. Issues can be used for feature requests but also for bugs or possible conflicts. The fashion in which these are taken care of shows a good picture on how serious the project is.
   \item \textit{Number of users} - The number of users influences the choice of packages in two ways. Firstly, if a lot of people use it, it is implicitly tested. So if there is a bug or vulnerability, it will eventually be found. The second reason is that if a vulnerability was to be found, a lot of people would experience this so it will be patched very fast.
   \item \textit{Backing company} - A backing company can ensure the project has enough resources to continue and therefore contributes to the stability of the project. However if such backing company gains too much influence, it can alter certain aspects of the project that would endanger its integrity.
   \item \textit{Ease of integration} - When a package is easy to integrate, it can be verified quickly and this is important when selecting packages.
   \item \textit{Community} - A large and active community contributes to the selection of a package since the power of open source is because of the diversity of contributors, users and documentation writers from different backgrounds to contribute to one project. 
   \item \textit{Known vulnerabilities} - The Common Vulnerabilities and Exposures (CVE) holds a database that has all known public security vulnerabilities. This database can be accessed to see if a project has had vulnerabilities. The number of vulnerabilities and the resolve time can provide a good indication of how serious the project is.
   \item \textit{Reputation} - The reputation can either be from the project itself, if it has made a name for itself. Or from the contributors or developing entity. One tends to believe that an entity with a good reputation creates good software, and that one with a bad reputation can create bad software. Yet an important aspect is that no reputation has no negative influence.
   \item \textit{Support} - Support is not a necessary aspect, however does positively influence the choice of a package if present. 
   \item \textit{Git stars} - The Git stars can give somewhat of an indication on the popularity of the project. 
   \item \textit{Knowledge of team} - Rather than an aspect about the project, this is an aspect of the current team a developer is operating in. To look at the strengths and weaknesses of the team, and start selecting packages with that in mind rather than choosing a package solely based on characteristics of that package.
     
\end{itemize}

\subsubsection{Economical factors}
Appendix \ref{ecoadoptionfactors} holds the found economical aspects that influence the decision of which package to choose. After every aspect there is at least one quote from experts that support this. The following economical aspects were found in the interviews that influenced the interviewees as follows:

\begin{itemize}
   \item \textit{License}  - Licenses may restrict the usage of projects for certain purposes and may influence one's own project. 
   \item \textit{Total cost of ownership} - The total costs are considered when selecting a package. When using open source software this is not very relevant most of the times. However this can sometimes result in some tricky business.
\end{itemize}

\subsection{Section 2, Trust factors and metrics}

The second section of the interview consists of two questions regarding trust in software. Trust factors as well as trust metrics were discussed in order to answer the following subquestion: Which factors contribute to gaining trust in a package?

\subsubsection{Trust factors}
For the first three categories, the interviewees will think of all the factors that influence the decision of what packages will be selected. After this, they will create a subset of all those factors that actually cause them to trust a certain package. While these two aspects are very related, they are definitely not the same. P1:\textit{ "For me the economical factors play a role in deciding what package, however not in trusting the package."}

As with the adoption factors, some participants named the actual factors where other named concrete examples of those factors. They are displayed separately to prevent information loss.
The found trust influencing aspects can also be categorized as technical and organizational aspects. Appendix \ref{techtrustfactors} holds the found technical trust influencing aspects. Each of these aspects is supported by at least one quote and appendix \ref{orgtrustfactors} holds the found organizational trust influencing aspects. Table \ref{trustfactorstable} displays all aspects that were named during the interviews:
\begin{table}[hb!]
\footnotesize
\resizebox{0.95\textwidth}{!}{\begin{minipage}{\textwidth}
\hskip-0.5cm\begin{tabular}{lllllllllllll}
\hline
 & \multicolumn{1}{c}{P1} & \multicolumn{1}{c}{P2} & \multicolumn{1}{c}{P3} & \multicolumn{1}{c}{P4} & \multicolumn{1}{c}{P5} & \multicolumn{1}{c}{P6} & \multicolumn{1}{c}{P7} & \multicolumn{1}{c}{P8} & \multicolumn{1}{c}{P9} & \multicolumn{1}{c}{P10} & \multicolumn{1}{c}{P11} & \multicolumn{1}{c}{P12} \\ \hline
\textbf{Technical} &  &  &  &  &  &  &  &  &  &  &  &  \\ \hline
\textit{Documentation} & \multicolumn{1}{c}{•} & \multicolumn{1}{c}{•} &  &  &  &  &  &  &  & \multicolumn{1}{c}{•} &  &  \\
\textit{Source code openness} & \multicolumn{1}{c}{•} & \multicolumn{1}{c}{•} &  &  &  &  &  &  &  &  &  & \multicolumn{1}{c}{•} \\
\textit{Code quality} &  & \multicolumn{1}{c}{•} &  &  &  &  & \multicolumn{1}{c}{•} &  & \multicolumn{1}{c}{•} &  &  &  \\ \hline
\textbf{Organizational} &  &  &  &  &  &  &  &  &  &  &  &  \\ \hline
\textit{Number of users} & \multicolumn{1}{c}{•} & \multicolumn{1}{c}{•} &  &  &  & \multicolumn{1}{c}{•} &  &  & \multicolumn{1}{c}{•} & \multicolumn{1}{c}{•} &  & \multicolumn{1}{c}{•} \\
\textit{Active maintenance} &  &  &  & \multicolumn{1}{c}{•} & \multicolumn{1}{c}{} & \multicolumn{1}{c}{•} &  &  &  &  &  &  \\
\textit{Community} &  &  & \multicolumn{1}{c}{•} &  &  & \multicolumn{1}{c}{•} & \multicolumn{1}{c}{•} &  &  &  & \multicolumn{1}{c}{•} & \multicolumn{1}{c}{•} \\
\textit{Contributors} &  &  &  &  & \multicolumn{1}{c}{•} &  & \multicolumn{1}{c}{•} & \multicolumn{1}{c}{•} & \multicolumn{1}{c}{•} &  &  &  \\
\textit{Git Issues} &  &  &  & \multicolumn{1}{c}{•} & \multicolumn{1}{c}{•} &  &  &  &  &  & \multicolumn{1}{c}{•} &  \\
\textit{Git stars} &  &  &  & \multicolumn{1}{c}{•} &  &  &  &  & \multicolumn{1}{c}{•} &  &  &  \\
\textit{Backing company} &  & \multicolumn{1}{c}{•} &  &  &  &  &  & \multicolumn{1}{c}{•} & \multicolumn{1}{c}{•} &  & \multicolumn{1}{c}{•} &  \\
\textit{Ease of integration} &  &  &  &  &  & \multicolumn{1}{c}{•} &  &  &  &  &  &  \\
\textit{Stack overflow activity} &  &  & \multicolumn{1}{c}{•} &  &  &  &  &  &  &  &  &  \\ \hline
\end{tabular}
\caption{Trust factors overview}
\label{trustfactorstable}
\end{minipage}}

\end{table}

There are some trust factors that are not mentioned at the adoption factors. These factors are:
\begin{itemize}
    \item \textit{Source code openness} - When the source code is available it provides the possibility to go through the actual code and to let one decide for itself whether or not this piece of code is trustworthy. However as P1 pointed out: "There are so many sneaky ways to get a backdoor into software..." This does not guarantee that the code is safe however does contribute to the possible ways of finding out. 
    \item \textit{Stackoverflow activity} - This is a factor that one participant named specifically, and it is related to the community and the issues. Since stackoverflow is a question-and-answer platform that displays the activeness of the community and users.
\end{itemize}

\subsubsection{Trust metrics}

Trust factors and trust metrics seem very alike, however there certainly is a distinction between them, moreover in the context that they were asked. One could argue that the trust metrics contain the concrete examples of trust factors, Which is true. However, the distinction for this research comes from the context in which the two were asked.

The question regarding the trust factors aims to uncover what aspects are relevant for an expert to gain trust in a software package. This question regards their personal perspective on what is important. Sometimes this would lead to concrete examples of trust factors: trust metrics. However, the question regarding trust metrics was asked after a detailed explanation about the TrustSECO project and what it aims to accomplish. A situation was described to the participants in which TrustSECO provided a trust score and they were asked what metrics they would value most in a trust score-breakdown. This contextual inconsistency makes all the difference here.

The remarkable aspect about this contextual difference is that some participants came up with metrics that represent a certain category. That were not named when asked about what factors are relevant when trusting a package, one question earlier.

As with the adoption and trust factors, some of the participants would name a category of metrics where others would name concrete examples. For now these will be named separately to prevent information loss. These can also be categorized in technical and organizational metrics.

Table \ref{trustmetricsoverview} shows what categories of metrics, or concrete examples of metrics were found:
\\
\begin{table}[h!]
\footnotesize
\resizebox{0.92\textwidth}{!}{\begin{minipage}{\textwidth}
\hskip-0cm\begin{tabular}{lllllllllllll}
\hline
 & \multicolumn{1}{c}{P1} & \multicolumn{1}{c}{P2} & \multicolumn{1}{c}{P3} & \multicolumn{1}{c}{P4} & \multicolumn{1}{c}{P5} & \multicolumn{1}{c}{P6} & \multicolumn{1}{c}{P7} & \multicolumn{1}{c}{P8} & \multicolumn{1}{c}{P9} & \multicolumn{1}{c}{P10} & \multicolumn{1}{c}{P11} & \multicolumn{1}{c}{P12} \\ \hline
\textbf{Technical} &  &  &  &  &  &  &  &  &  &  &  &  \\ \hline
\textit{Code quality} &  & \multicolumn{1}{c}{•} &  &  &  &  &  & \multicolumn{1}{c}{•} & \multicolumn{1}{c}{•} &  &  & \multicolumn{1}{c}{•} \\
\textit{Complexity} &  &  &  & \multicolumn{1}{c}{•} &  &  &  & \multicolumn{1}{c}{•} &  &  &  &  \\
\textit{Tests} &  & \multicolumn{1}{c}{•} &  & \multicolumn{1}{c}{•} &  & \multicolumn{1}{c}{•} &  &  &  &  &  &  \\ \hline
\textbf{Organizational} &  &  &  &  &  &  &  &  &  &  &  &  \\ \hline
\textit{Active maintenance} &  &  & \multicolumn{1}{c}{•} &  &  & \multicolumn{1}{c}{•} & \multicolumn{1}{c}{•} &  & \multicolumn{1}{c}{•} & \multicolumn{1}{c}{•} &  &  \\
\textit{Number of contributors} &  &  &  &  &  & \multicolumn{1}{c}{•} &  & \multicolumn{1}{c}{•} & \multicolumn{1}{c}{•} &  &  &  \\
\textit{Users} &  & \multicolumn{1}{c}{•} & \multicolumn{1}{c}{•} & \multicolumn{1}{c}{•} &  & \multicolumn{1}{c}{•} &  & \multicolumn{1}{c}{•} & \multicolumn{1}{c}{•} & \multicolumn{1}{c}{•} &  &  \\
\textit{Stability metrics} &  &  &  &  & \multicolumn{1}{c}{•} &  &  &  &  &  & \multicolumn{1}{c}{•} &  \\
\textit{Supporting platforms} &  &  &  &  &  &  & \multicolumn{1}{c}{•} &  &  &  &  &  \\
\textit{Reputation} &  &\multicolumn{1}{c}{•}  &  &  &  &  &  &  &  &  & \multicolumn{1}{c}{•} &  \\
\textit{Git issues} &  &  &  &  & \multicolumn{1}{c}{•} &  & \multicolumn{1}{c}{•} &  & \multicolumn{1}{c}{•} &  &  & \multicolumn{1}{c}{•} \\
\textit{Known vulnerabilities} &  &  & \multicolumn{1}{c}{•} &  & \multicolumn{1}{c}{•} &  & \multicolumn{1}{c}{•} &  &  &  &  & \multicolumn{1}{c}{•} \\
\textit{Stack overflow} &  &  &  &  &  &  & \multicolumn{1}{c}{•} &  &  &  &  &  \\
\textit{Developing entity} &  & \multicolumn{1}{c}{•} &  &  &  &  &  &  &  & \multicolumn{1}{c}{•} &  &  \\ \hline
\end{tabular}
\caption{Trust metrics overview}
\label{trustmetricsoverview}
\end{minipage}}
\end{table}

One  may notice that P1 does not have any trust metrics, this is because the question was added after the first interview.
Appendix \ref{techtrustmetrics} and \ref{orgtrustmetrics} hold a list of all found categories of metrics and metrics themselves with quotes to back them up.

\subsection{Section 3, Personal bias and experience }

This section of the interview consists of three questions aimed to answer the question: How do personal aspects influence trust in a package?
\\ \\
This section was created since the literature mainly concerned characteristics of the packages and projects, rather than the actual people who have to make decisions based on these characteristics. The difficult part in this is that there is no one correct answer as to how personal aspects influence this trust. This is because each of the participants has another perspective on this due to different past experiences. However there is one main conclusion to be drawn from this. That is that each of the participants have changed the way they look at selecting software over the years. This is due to the experiences they, but also experiences that the people around them had. Some have experienced more severe situations than others, and this results in them being more careful and paying more attention to this problem. 

Another important aspect that could be concluded from this last section of the interview, is that some participants came up with additional factors or metrics that were not mentioned before. This illustrates that this is not a subject that is consciously thought about a lot, since it takes some thought process to come up with all the relevant aspects related to this matter. Hence proving the importance of this research. This then also implicitly means that there could be more relevant factors that are not uncovered during the interviews and would be if there would be more or different questions. Some participants, who had taken prior interest in this matter did not experience this growth in the interview since they had given this subject some more in-depth thought at some point in their career.

The interviews lead to the thought that there is no one-size-fits-all list of factors why a certain individual trusts a certain package since it there was a lot of variance between participants. However that by looking at enough individuals, a well defined list of factors can be defined that determines if a package is trustworthy. \\

\newpage
\chapter{Discussion}
One of the most prominent differences between the results from the literature and interviews, is the presence of technical factors for both the adoption and trust aspects. Technical factors are a lot more represented in literature than in the interviews. This is because only the factors or category of factors was noted if the participant named it before they were asked about it explicitly. Which revealed the following: a lot of the technical factors that were named in literature have to hold for interviewees to even consider looking at other factors. So this is definitely relevant when either adopting or trusting a package, and has to match with the current project in order for the project to be looked further into. 
\\ \\
Another very important difference between some results are the results between the trust factors and trust metrics. First the participants were what factors could influence the trust on a package. After this they were asked what metrics they would like to seen in an overview for determining trust. The question was initially not for this research and was to get a better idea for the general TrustSECO project, however turned out to be quite the contrary. These two questions would sometimes give vastly different answers.

This is due to a complication that this research has seen previously. Namely the difference in naming categories of factors vs. naming concrete examples of factors and the fact that there is a form of growth in the matter as the interview progresses. When discussing the trust metrics, the participants would sometimes name completely different aspects since the aspects had to be somewhat measurable and had not thought of certain aspects before. 

This then also illustrates that there could be more trust factors than the participants have named collectively, since a lot of the steps and thought processes they go through are implicit.
\subsubsection{TrustSECO}
This research set out to create a list of adoption and trust factors for software packages. The purpose of this list is to serve as guidance for creating a survey that will sort the factors on importance. This will then be a major guidance for the TrustSECO project to weigh certain metrics connected to the prioritized factors, in order to create the final project. So from now on a survey needs to be created that ranks all the found factors, as well as all the metrics. The metrics are eventually going to be the most relevant for the TrustSECO project. A prototype of the project was recently created during Odyssey Momentum. This is a mass-online collaboration event in which the whole TrustSECO team invested a whole weekend to start creating the first version of the software. Appendix 7.6 holds more detail regarding my personal experience for this event. 

\chapter{Conclusion}
This research has shown that there are a lot of different factors that play a role when selecting or trusting software packages. It has also shown that these factors have a different impact on stakeholders depending on their role and the and situation. 

For this research, there is a distinction between adoption factors and trust factors, both in literature and in the interviews. The adoption factors in literature share a great similarity with the ones found in the interviews. There are some differences. For example the interviews hold concrete examples of the better formulated categories in literature and there are few aspects that really differ from each other. Where the trust factors found in literature differ vastly from the majority of the found trust factors in the interviews. In literature the technical factors are the highest scoring factors where in the interviews the organizational seemed to be the most important.

Then there is the comparison between all found adoption and trust factors. Where the adoption factors display all the factors that are looked upon when selecting a package, the trust factors aim to illustrate what factors make people gain trust in a package. For the latter, this research has shown the difficulty to create a one size fits all trust factor list for individuals to gain trust in a package. This is because each individual has his or hers own past experiences with this matter, and thus has a different perspective on this. However, a good estimation of factors that make a package trustworthy can be generated by learning from enough individuals' experiences.

\newpage
\chapter{Appendix}
\section{Literature review}
\subsection{Article list SLR}
\begin{table}[hb!]
\resizebox{0.85\textwidth}{!}{\begin{minipage}{\textwidth}
\hskip-2.0cm\begin{tabular}{|l|l|l|}
\hline
\textbf{Article title} & \textbf{Author} & \textbf{Year} \\ \hline
The infeasibility of experimental quantification of life-critical software reliability & RW Butler, GB Finelli & 1991 \\ \hline
The infeasibility of quantifying the reliability of life-critical real-time software & RW Butler, GB Finelli & 1993 \\ \hline
Software reliability and system reliability & JC Laprie, K Kanoun & 1996 \\ \hline
Method and system for determining software reliability & MR Siegel, JI Ferrell & 1996 \\ \hline
\begin{tabular}[c]{@{}l@{}}Predicting software reliability from testing taking into account other knowledge about a \\  program\end{tabular} & A Bertolino, L Strigini & 1996 \\ \hline
Understanding the sources of variation in software inspections & \begin{tabular}[c]{@{}l@{}}A Porter, H Siy, A Mockus,\\  L Votta\end{tabular} & 1998 \\ \hline
Software metrics: successes, failures and new directions & NE Fenton, M Neil & 1999 \\ \hline
The paradoxes of free software & SM McJohn & 2000 \\ \hline
\begin{tabular}[c]{@{}l@{}}Open source software projects as virtual organisations: competency rallying for \\ software development\end{tabular} & K Crowston, B Scozzi & 2002 \\ \hline
\begin{tabular}[c]{@{}l@{}}Government preferences for promoting open-source software: A solution in search\\  of a problem\end{tabular} & B Reddy, DS Evans & 2002 \\ \hline
\begin{tabular}[c]{@{}l@{}}Why hackers do what they do: Understanding motivation and effort in free/open\\  source software projects\end{tabular} & KR Lakhani, RG Wolf & 2003 \\ \hline
\begin{tabular}[c]{@{}l@{}}Motivation of software developers in Open Source projects: an Internet-based \\ survey of contributors to the Linux kernel\end{tabular} & \begin{tabular}[c]{@{}l@{}}G Hertel, S Niedner,\\  S Herrmann\end{tabular} & 2003 \\ \hline
Why open source software can succeed & A Bonaccorsi, C Rossi & 2003 \\ \hline
\begin{tabular}[c]{@{}l@{}}Open-source software development as gift culture: Work and identity formation in\\  an internet community\end{tabular} & M Bergquest & 2003 \\ \hline
Open source software for the public administration & \begin{tabular}[c]{@{}l@{}}GL Kovács, S Drozdik,\\  P Zuliani…\end{tabular} & 2004 \\ \hline
Open source software and open data standards in public administration & \begin{tabular}[c]{@{}l@{}}GL Kovács, S Drozdik, \\ P Zuliani…\end{tabular} & 2004 \\ \hline
The Collaborative Integrity of Open-Source Software & GR Vetter & 2004 \\ \hline
Resistance as motivation for innovation: Open source software & JF Kavanagh & 2004 \\ \hline
Agents of responsibility in software vulnerability processes & \begin{tabular}[c]{@{}l@{}}A Takanen, P Vuorijärvi,\\  M Laakso, J Röning\end{tabular} & 2004 \\ \hline
\end{tabular}
\end{minipage}}
\label{SLR}
\end{table}
\newpage

\begin{table}[ht!]
\resizebox{0.8\textwidth}{!}{\begin{minipage}{\textwidth}
\hskip-2.0cm\begin{tabular}{|l|l|l|}
\hline
\textbf{Article title} & \textbf{Author} & \textbf{Year} \\ \hline
\begin{tabular}[c]{@{}l@{}}Relationships between open source software companies and communities: \\ Observations from Nordic firms\end{tabular} & L Dahlander, MG Magnusson & 2005 \\ \hline
Participant satisfaction with open source software & BL Chawner & 2005 \\ \hline
\begin{tabular}[c]{@{}l@{}}Motivation, governance, and the viability of hybrid forms in open source software\\  development\end{tabular} & SK Shah & 2006 \\ \hline
\begin{tabular}[c]{@{}l@{}}Assessing the Impact of Project Founder Reputation and Project Structure on\\  Motivation to Participate in Open Source Software Projects\end{tabular} & \begin{tabular}[c]{@{}l@{}}K Ghosh, J Ziegelmayer, \\ A Ammeter\end{tabular} & 2006 \\ \hline
\begin{tabular}[c]{@{}l@{}}Location, location, location: How network embeddedness affects project success\\  in open source systems\end{tabular} & \begin{tabular}[c]{@{}l@{}}R Grewal, GL Lilien,\\  G Mallapragada\end{tabular} & 2006 \\ \hline
\begin{tabular}[c]{@{}l@{}}Impacts of license choice and organizational sponsorship on user interest and\\  development activity in open source software projects\end{tabular} & KJ Stewart, AP Ammeter… & 2006 \\ \hline
Software estimation: demystifying the black art & S McConnell & 2006 \\ \hline
Bounty programs in free/libre/open source software & S Krishnamurthy, AK Tripathi & 2006 \\ \hline
A software component quality model: A preliminary evaluation & A Alvaro, ES De Almeida… & 2006 \\ \hline
OSS opportunities in open source software—CRM and OSS standards & \begin{tabular}[c]{@{}l@{}}G Bruce, P Robson, \\ R Spaven\end{tabular} & 2006 \\ \hline
\begin{tabular}[c]{@{}l@{}}New Perspectives on Public Goods Production: Policy Implications of Open Source\\  Software\end{tabular} & JA Lee & 2006 \\ \hline
\begin{tabular}[c]{@{}l@{}}Developing an open source software development process model using grounded\\  theory\end{tabular} & Y Tian & 2006 \\ \hline
A Reputation-Based Mechanism for Software Vulnerability Disclosure & X Zhao & 2007 \\ \hline
\begin{tabular}[c]{@{}l@{}}The governance of free/open source software projects: monolithic, multidimensional,\\  or configurational?\end{tabular} & ML Markus & 2007 \\ \hline
Intrinsic motivation in open source software development & \begin{tabular}[c]{@{}l@{}}J Bitzer, W Schrettl, \\ PJH Schröder\end{tabular} & 2007 \\ \hline
\begin{tabular}[c]{@{}l@{}}An empirical analysis of the impact of software vulnerability announcements on\\  firm stock price\end{tabular} & R Telang, S Wattal & 2007 \\ \hline
Reputation in Open Source Software Virtual Communities & \begin{tabular}[c]{@{}l@{}}LV Casaló, J Cisneros,\\  C Flavián…\end{tabular} & 2008 \\ \hline
\begin{tabular}[c]{@{}l@{}}Emergence of new project teams from open source software developer networks:\\  Impact of prior collaboration ties\end{tabular} & \begin{tabular}[c]{@{}l@{}}J Hahn, JY Moon, \\ C Zhang\end{tabular} & 2008 \\ \hline
Temporal metrics for software vulnerabilities & \begin{tabular}[c]{@{}l@{}}JA Wang, F Zhang, \\ M Xia\end{tabular} & 2008 \\ \hline
Method and apparatus for detecting vulnerabilities and bugs in software applications & \begin{tabular}[c]{@{}l@{}}VC Sreedhar, GF Cretu,\\  JT Dolby\end{tabular} & 2008 \\ \hline
\begin{tabular}[c]{@{}l@{}}User and developer mediation in an Open Source Software community: Boundary\\  spanning through cross participation in online discussions\end{tabular} & \begin{tabular}[c]{@{}l@{}}F Barcellini, F Détienne,\\  JM Burkhardt\end{tabular} & 2008 \\ \hline
An approach for selecting software-as-a-service (SaaS) product & M Godse, S Mulik & 2009 \\ \hline
Impact of license choice on open source software development activity & J Colazo, Y Fang & 2009 \\ \hline
Research on testing-based software credibility measurement and assessment & Q Hongbing, Z Xiaojie… & 2009 \\ \hline
\begin{tabular}[c]{@{}l@{}}Designers wanted: participation and the user experience in open source software\\  development\end{tabular} & \begin{tabular}[c]{@{}l@{}}PM Bach, R DeLine, \\ JM Carroll\end{tabular} & 2009 \\ \hline
3.5 Open Source Software Research and Blockchain & J Lindman & 2009 \\ \hline
“Constructing the users” in open source software development & N Iivari & 2009 \\ \hline
System and method for maximizing software package license utilization & \begin{tabular}[c]{@{}l@{}}S Varadarajan, G Sridhar,\\  KK Rao\end{tabular} & 2010 \\ \hline
Software metrics and software metrology & A Abran & 2010 \\ \hline
\begin{tabular}[c]{@{}l@{}}Creating and evolving developer documentation: understanding the decisions of\\  open source contributors\end{tabular} & B Dagenais, MP Robillard & 2010 \\ \hline
Code forking in open-source software: a requirements perspective & \begin{tabular}[c]{@{}l@{}}NA Ernst, S Easterbrook, \\ J Mylopoulos\end{tabular} & 2010 \\ \hline
Trust and reputation for successful software self-organisation & JM Seigneur, P Dondio & 2011 \\ \hline

\end{tabular}
\end{minipage}}
\end{table}

\begin{table}[h!]
\resizebox{0.77\textwidth}{!}{\begin{minipage}{\textwidth}
\hskip-2.0cm\begin{tabular}{|l|l|l|}
\hline
  
\textbf{Article title} & \textbf{Author} & \multicolumn{1}{l|}{\textbf{Year}} \\ \hline
  
\begin{tabular}[c]{@{}l@{}}SLA-based resource allocation for software as a service provider (SaaS) in cloud \\ computing environments\end{tabular} & L Wu, SK Garg, R Buyya & 2011 \\ \hline
  
\begin{tabular}[c]{@{}l@{}}A systematic literature review on fault prediction performance in software\\  engineering\end{tabular} & \begin{tabular}[c]{@{}l@{}}T Hall, S Beecham, \\ D Bowes, D Gray…\end{tabular} & 2011 \\ \hline
  
Software quality: theory and management & A Gillies & 2011 \\ \hline
  
\begin{tabular}[c]{@{}l@{}}A risk assessment framework for evaluating Software-as-a-Service (SaaS) cloud\\  services before adoption\end{tabular} & L Bernard & 2011 \\ \hline
  
Understanding broadcast based peer review on open source software projects & PC Rigby, MA Storey & 2011 \\ \hline
  
A theory-grounded framework of Open Source Software adoption in SMEs & RD Macredie, K Mijinyawa & 2011 \\ \hline
  
\begin{tabular}[c]{@{}l@{}}Understanding open source software peer review: Review processes, parameters\\  and statistical models, and underlying behaviours and mechanisms\end{tabular} & PC Rigby & 2011 \\ \hline
  
\begin{tabular}[c]{@{}l@{}}Design and evaluation of a process for identifying architecture patterns in open\\  source software\end{tabular} & \begin{tabular}[c]{@{}l@{}}KJ Stol, P Avgeriou,\\  MA Babar\end{tabular} & 2011 \\ \hline
  
\begin{tabular}[c]{@{}l@{}}Analyzing and Identifying SaaS for Development of a Project by calculating its\\  Reputation\end{tabular} & BR Rao & 2012 \\ \hline
  
\begin{tabular}[c]{@{}l@{}}Carrots and rainbows: Motivation and social practice in open source software\\  development\end{tabular} & \begin{tabular}[c]{@{}l@{}}G Von Krogh, S Haefliger,\\  S Spaeth, MW Wallin\end{tabular} & 2012 \\ \hline
  
A model of open source developer foundations & D Riehle, S Berschneider & 2012 \\ \hline
  
Research of trustworthy software system in the network & \begin{tabular}[c]{@{}l@{}}Y Liu, L Zhang, \\ P Luo, Y Yao\end{tabular} & 2012 \\ \hline
  
\begin{tabular}[c]{@{}l@{}}Study on credibility level of trustworthy software development process based\\  on grey nonlinear cluster\end{tabular} & \begin{tabular}[c]{@{}l@{}}S Liu, J Forrest, Y \\ Yangjie, K Zhang, \\ C Mi, N Xie…\end{tabular} & 2012 \\ \hline
  
A non-functional requirements tradeoff model in trustworthy software & \begin{tabular}[c]{@{}l@{}}MX Zhu, XX Luo,\\  XH Chen, DD Wu\end{tabular} & 2012 \\ \hline
  
How peripheral developers contribute to open-source software development & P Setia, B Rajagopalan… & 2012 \\ \hline
  
\begin{tabular}[c]{@{}l@{}}Research note—Lock-in strategy in software competition: Open-source software\\  vs. proprietary software\end{tabular} & KX Zhu, ZZ Zhou & 2012 \\ \hline
  
Why do commercial companies contribute to open source software? & \begin{tabular}[c]{@{}l@{}}M Andersen-Gott, \\ G Ghinea, B Bygstad\end{tabular} & 2012 \\ \hline
  
\begin{tabular}[c]{@{}l@{}}Do the allocation and quality of intellectual assets affect the reputation of open\\  source software projects?\end{tabular} & R Méndez-Durón & 2013 \\ \hline
  
Towards reputation-as-a-service & C Hillebrand, M Coetzee & 2013 \\ \hline
  
Software fault prediction metrics: A systematic literature review & \begin{tabular}[c]{@{}l@{}}D Radjenović, \\ M Heričko, R Torkar…\end{tabular} & 2013 \\ \hline
  
Automatic polymorphic exploit generation for software vulnerabilities & \begin{tabular}[c]{@{}l@{}}M Wang, P Su, Q Li,\\  L Ying, Y Yang, D Feng\end{tabular} & 2013 \\ \hline
  
'Computing'Requirements in Open Source Software Projects & \begin{tabular}[c]{@{}l@{}}X Xiao, A Lindberg,\\  S Hansen, K Lyytinen\end{tabular} & 2013 \\ \hline
  
\begin{tabular}[c]{@{}l@{}}From closed to open: Job role changes, individual predispositions, and the adoption\\  of commercial open source software development\end{tabular} & \begin{tabular}[c]{@{}l@{}}O Alexy, J Henkel,\\  MW Wallin\end{tabular} & 2013 \\ \hline
  
Learning and best practices for learning in open-source software communities & V Singh, L Holt & 2013 \\ \hline
  
\begin{tabular}[c]{@{}l@{}}All complaints are not created equal: text analysis of open source software defect \\ reports\end{tabular} & U Raja & 2013 \\ \hline
  
\begin{tabular}[c]{@{}l@{}}How social QnA sites are changing knowledge sharing in open source software \\ communities\end{tabular} & \begin{tabular}[c]{@{}l@{}}B Vasilescu, A Serebrenik,\\  P Devanbu…\end{tabular} & 2014 \\ \hline
  
\begin{tabular}[c]{@{}l@{}}Secured trust and reputation system: analysis of malicious behaviors and \\ optimization\end{tabular} & A Bradai & 2014 \\ \hline
  
\begin{tabular}[c]{@{}l@{}}Measuring the health of open source software ecosystems: Beyond the scope of\\  project health\end{tabular} & S Jansen & 2014 \\ \hline
  
Software Reliability: State of the Art Report 14: 2 & A Bendell, P Mellor & 2014 \\ \hline
  
Auditing and maintaining provenance in software packages & Q Pham, T Malik, I Foster & 2014 \\ \hline
  
\begin{tabular}[c]{@{}l@{}}Estimating development effort in free/open source software projects by mining\\  software repositories: a case study of openstack\end{tabular} & \begin{tabular}[c]{@{}l@{}}G Robles, JM Gonzále\\ -Barahona, C Cervigón…\end{tabular} & 2014 \\ \hline
\end{tabular}
\end{minipage}}
\end{table}
\newpage
\begin{table}[ht!]
\resizebox{0.8\textwidth}{!}{\begin{minipage}{\textwidth}
\hskip-2.0cm\begin{tabular}{|l|l|l|}
\hline
\textbf{Article title} & \textbf{Author} & \textbf{Year} \\ \hline
\begin{tabular}[c]{@{}l@{}}Transactive memory system, communication quality, and knowledge sharing in distributed\\  teams: An empirical examination in open source software project teams\end{tabular} & X Chen & 2014 \\ \hline
The Spack package manager: bringing order to HPC software chaos & \begin{tabular}[c]{@{}l@{}}T Gamblin, M LeGendre,\\  MR Collette…\end{tabular} & 2015 \\ \hline
Analysis and assessment of software library projects & \begin{tabular}[c]{@{}l@{}}JW Nicol, BL Roberts, \\ JO Pillgram-Larsen…\end{tabular} & 2015 \\ \hline
Software applications have on average 24 vulnerabilities inherited from buggy components & L Constantin & 2015 \\ \hline
\begin{tabular}[c]{@{}l@{}}Raising the general public's awareness and adoption of open source software through social\\  QnA interactions\end{tabular} & N Choi, K Yi & 2015 \\ \hline
Group Reputation in an Open Source Software Community: Antecedents and Outcomes & Y Cai, D Zhu & 2016 \\ \hline
Maintenance effort estimation for open source software: A systematic literature review & \begin{tabular}[c]{@{}l@{}}H Wu, L Shi,\\  C Chen, Q Wang…\end{tabular} & 2016 \\ \hline
Modeling library dependencies and updates in large software repository universes & \begin{tabular}[c]{@{}l@{}}RG Kula, C De Roover, \\ DM German, T Ishio…\end{tabular} & 2017 \\ \hline
Secure dependency enforcement in package management systems & \begin{tabular}[c]{@{}l@{}}L Catuogno, C Galdi, \\ G Persiano\end{tabular} & 2017 \\ \hline
Large-scale Modeling, Analysis, and Preservation of Free and Open Source Software & S Zacchiroli & 2017 \\ \hline
Open Source Software Hosting Platforms: A Collaborative Perspective's Review. & G Alamer, S Alyahya & 2017 \\ \hline
Software processes analysis with provenance & \begin{tabular}[c]{@{}l@{}}GCB Costa, HLO Dalpra,\\  EN Teixeira…\end{tabular} & 2018 \\ \hline
Software Provenance: Track the Reality Not the Virtual Machine & \begin{tabular}[c]{@{}l@{}}D Wilkinson, L Oliveira,\\  D Mossé…\end{tabular} & 2018 \\ \hline
Hackers vs. testers: A comparison of software vulnerability discovery processes & \begin{tabular}[c]{@{}l@{}}D Votipka, R Stevens,\\  E Redmiles, J Hu…\end{tabular} & 2018 \\ \hline
A business model for commercial open source software: A systematic literature review & \begin{tabular}[c]{@{}l@{}}S Shahrivar, S Elahi,\\  A Hassanzadeh…\end{tabular} & 2018 \\ \hline
Collaborative SLA and reputation-based trust management in cloud federations & \begin{tabular}[c]{@{}l@{}}K Papadakis-\\ Vlachopapadopoulos…\end{tabular} & 2019 \\ \hline
A systematic examination of knowledge loss in open source software projects & \begin{tabular}[c]{@{}l@{}}M Rashid, PM Clarke,\\  RV O'Connor\end{tabular} & 2019 \\ \hline
\begin{tabular}[c]{@{}l@{}}THE TAKEOFF OF OPEN SOURCE SOFTWARE: A SIGNALING PERSPECTIVE\\  BASED ON COMMUNITY ACTIVITIES.\end{tabular} & \begin{tabular}[c]{@{}l@{}}P Setia, BL Bayus,\\  B Rajagopalan\end{tabular} & 2020 \\ \hline
\end{tabular}
\end{minipage}}
\end{table}

\newpage

\newpage
\section{Interviews}
\subsection{Informed consent}
\begin{figure}[hb!]
    \centering
    \hspace*{-1.0cm}
    \includegraphics[scale=0.57]{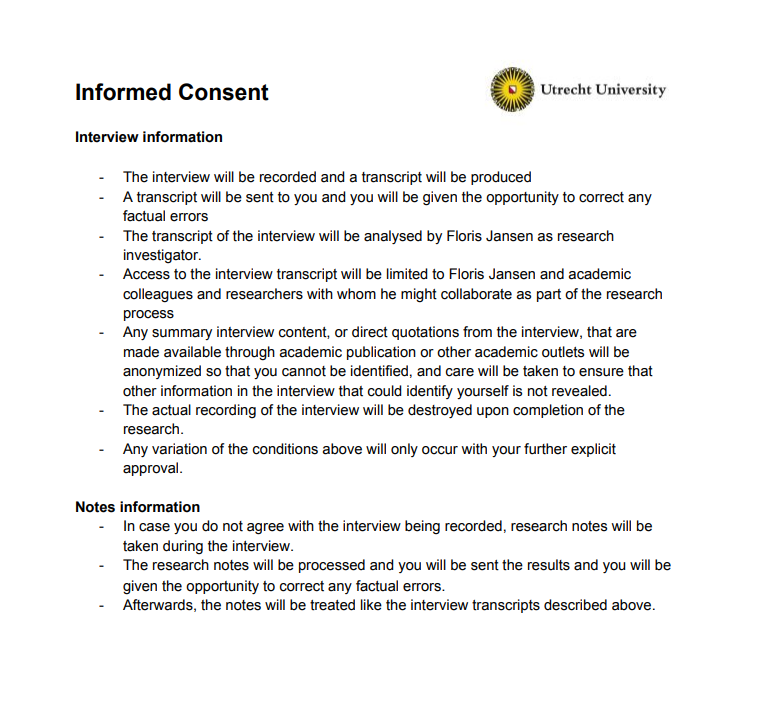}
\end{figure}
\label{informedconsent}
\newpage

\begin{figure}[h!]
    \centering
    \hspace*{-1.0cm}
    \includegraphics[scale=0.86]{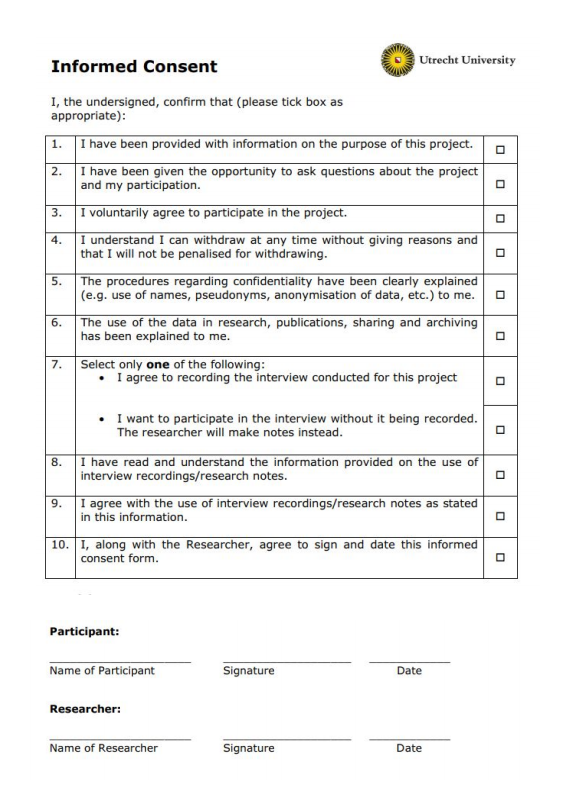}
\end{figure}
\newpage
\newpage

\subsection{Interview layout}
\begin{figure}[htbp]
\centering
\hspace*{-1.8 in}
    \includegraphics[scale=0.252]{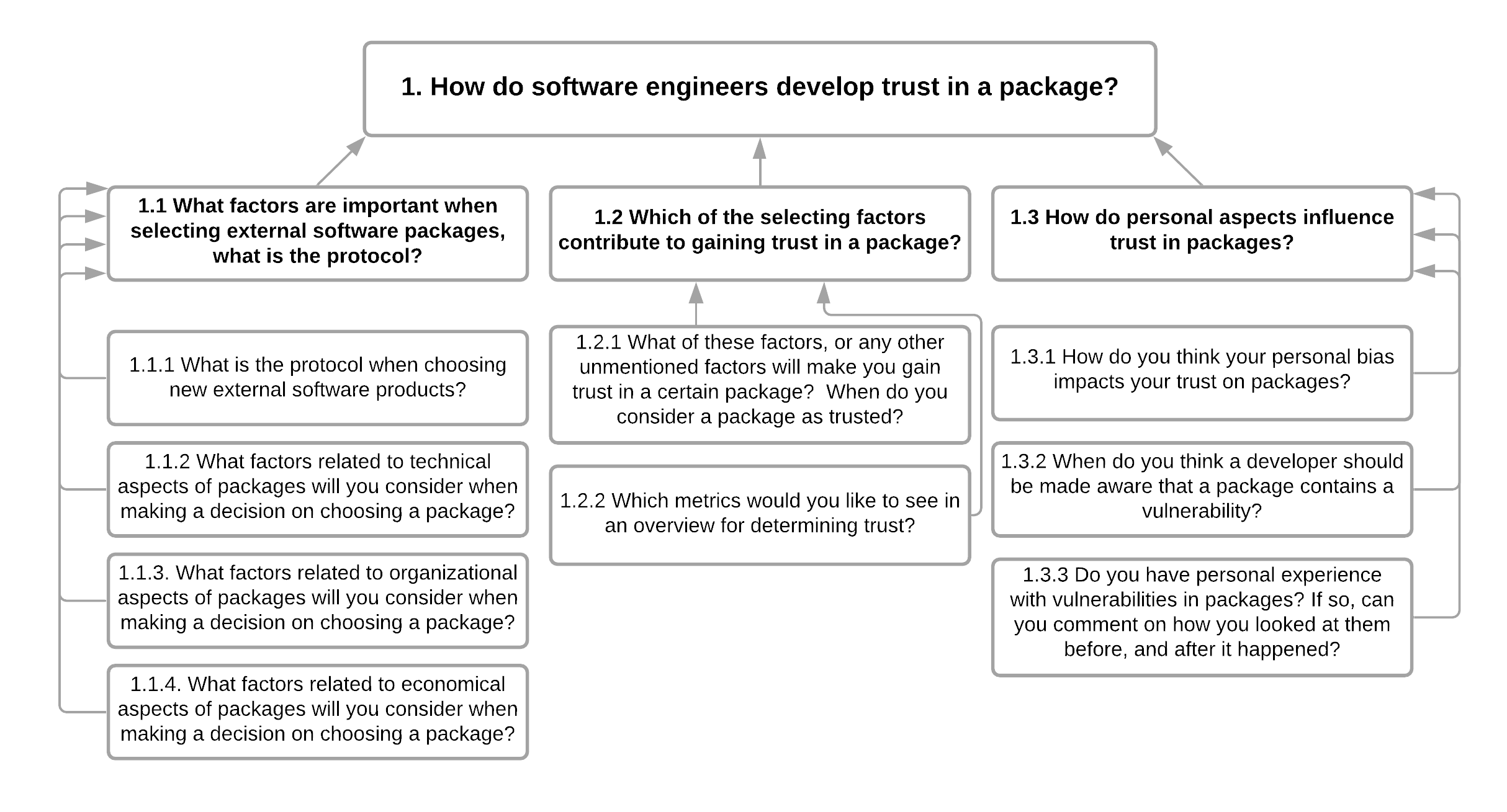}
    \label{interviewlayout}
\end{figure}

\newpage
\subsection{Interview protocol}
\begin{table}[ht!]
\resizebox{1.0\textwidth}{!}{\begin{minipage}{\textwidth}
\hskip-2.0cm\begin{tabular}{|l|}
\hline
\begin{tabular}[c]{@{}l@{}}Utrecht University is researching the factors that influence software packages selection. Such an \\ impact factor can be defined as a characteristic of a software package that results in trust in that \\ package. The results of these interviews, together with a literature study will lead to a survey that, \\ if done on a large scale, will reveal what these impact factors are. This interview will be about \\ 45 minutes. This information will be available through the informed-consent, that needs to be \\ signed before the interview starts.\end{tabular} \\ \hline
 \\ \hline
\begin{tabular}[c]{@{}l@{}}State: name, age, nationality, education, function(profession), years of experience, organization, \\ involvement in choosing packages, selection process used in the organization, type of organization,\\  industrial sector of organization, usage of external products within organization\end{tabular} \\ \hline
 \\ \hline
1. What is the protocol when choosing new external software products? \\ \hline
 \\ \hline
\begin{tabular}[c]{@{}l@{}}2. What factors related to technical aspects of the packages will you consider when making a \\ decision on choosing a package?\end{tabular} \\ \hline
 \\ \hline
\begin{tabular}[c]{@{}l@{}}3. What factors related to organizational aspects of the packages will you consider when making\\  a decision on choosing a package?\end{tabular} \\ \hline
 \\ \hline
\begin{tabular}[c]{@{}l@{}}4. What factors related to economical aspects of the packages will you consider when making a\\  decision on choosing a package?\end{tabular} \\ \hline
 \\ \hline
\begin{tabular}[c]{@{}l@{}}5. What of these factors, or any other unmentioned factors will make you gain trust in a package, \\ when do you consider a  package trusted?\end{tabular} \\ \hline
 \\ \hline
6. What metrics would you like to see in an overview for determining trust? \\ \hline
 \\ \hline
7. How do you think your personal bias impacts your trust on packages? \\ \hline
 \\ \hline
8. When do you think a developer should be made aware that a package contains a vulnerability? \\ \hline
 \\ \hline
\begin{tabular}[c]{@{}l@{}}9. Do you have personal experience with vulnerabilities in packages? If so, can you comment on \\ how you looked at them before, and after it happened?\end{tabular} \\ \hline
\end{tabular}
\end{minipage}}
\label{interviewprotocol}
\end{table}

\newpage
\section{Adoption factors}
    \subsection{Technical adoption factors}
\begin{table}[ht!]
\centering
\resizebox{0.81\textwidth}{!}{\begin{minipage}{\textwidth}
\hskip-2.0cm\begin{tabular}{|l|}
\hline
\textbf{Compatibility} \\ \hline
• "Well first of all it has to fit within the current application" \\
• "So the most important one is that it is compliant with your current framework." \\
• "First we take a look how secure the project is, and how compliant it is with our current build" \\
• "I globally scan the code to see if I understand what it does, also to see if it fits my use case" \\
• "We take a look at the functionality, and does this fit within the current framework" \\
• "I look at the documentation to see what datastructures are used, and if they are compatible with \\ my project"\\ \hline
\textbf{Documentation} \\ \hline
• "Good comments and a good readme are crucial" \\
• "With Python I mainly look at the documentation, what does it look like ..." \\
• "dependant on the size and complexity, the documentation can be really important or irrelevant..." \\
\begin{tabular}[c]{@{}l@{}}• "If every new function requires 2 days of reverse engineering because the documentation is bad, \\ drop this package!"\end{tabular} \\
\begin{tabular}[c]{@{}l@{}}• "if there is no documentation I have to figure it out myself. It makes it hard for us to work with so\\  it is not a matter of trust but then it comes back down to the pillar's ease of use."\end{tabular} \\
• "... and after that you look at things like documentation and community" \\
\begin{tabular}[c]{@{}l@{}}• "Documentation can be a good indication on how mature a project is, and that multiple people are\\  seriously working on this project."\end{tabular} \\
• "By looking at the documentation we can see if it is a good or bad package" \\
• "I check that by looking at the documentation or the API" \\ \hline
\textbf{Complexity} \\ \hline
\begin{tabular}[c]{@{}l@{}}• "If I want to solve a problem, I want to find a dependency that only solves this problem and \\ nothing else"\end{tabular} \\
\begin{tabular}[c]{@{}l@{}}• "It should be compact, especially with Python  people do this elegantly, is should serve a \\ single purpose and be clear"\end{tabular} \\
• "A solution should not be too complex for what it is trying to solve, it should make the coding easier" \\
\begin{tabular}[c]{@{}l@{}}• "Oh I forgot to mention, I also check some of the source code of the project. To get a feel on how the \\ comments are, how complex are the functions etc. So if i want to make little adjustments I can easily \\ understand how and where this should be done."\end{tabular} \\
• "...  if that is the case we might use it, one of the aspects we then look at is the size of the package" \\
• "One other aspect to look at is:  how difficult is it to replace that component? \\ \hline
\textbf{Security} \\ \hline
• "First we take a look how secure the project is, and how compliant it is with our current build" \\
• "We take a look at how easy it is to integrate in our system, and whether or not it is secure" \\
\begin{tabular}[c]{@{}l@{}}• "We think security is very important, so we use several static code analysis tools created \\ by the software improvement group"\end{tabular} \\ \hline
\textbf{Source code  openness} \\ \hline
• "I globally scan the code to see if I understand what it does, also to see if it fits my use case" \\
• "If you can read what the code does exactly, that is very important  for me" \\ \hline
\textbf{Usability} \\ \hline
• "And releases are very important, and also the usability aspect should not be overlooked" \\ \hline
\textbf{Customization} \\ \hline
\begin{tabular}[c]{@{}l@{}}• "I want a package that is not too complex for what it should do, so you want it to be \\ customizable enough to ensure it can be used for your purpose."\end{tabular} \\ \hline
\textbf{Code Quality} \\ \hline
• "If there are tests available, so you can see that some functions really do the things they claim they do" \\
• "There are also several code quality tools that grade a package, they are also good indicators to watch" \\
\begin{tabular}[c]{@{}l@{}}• "It should do what you need it do you, and what it says that it does. If that is not the case then \\ I will not even consider it"\end{tabular} \\ \hline
\end{tabular}
\end{minipage}}
\label{techadoptionfactors}
\end{table}

\newpage
    \subsection{Organizational adoption factors}

\begin{table}[ht!]
\resizebox{0.78\textwidth}{!}{\begin{minipage}{\textwidth}
\hskip-1.2cm\begin{tabular}{|l|}
\hline
\textbf{Active maintenance} \\ \hline
• "For me one of the first things I look at it how regularly it is updated" \\
"When was it updated for the last time, first time in 3 years or recently?" \\
• "If the last update was in 2016, we will not even consider this package" \\
• "You could then use an external library, yet you want to know how active it is and if  it is maintained well " \\
• "The first step for me is to go to GitHub and check how often there are new commits" \\
• "the second step is how regular those commits are..."" \\
• "An important factor is to get the feeling that the project is actively maintained, so check the list of releases" \\
• "They are constantly updating and releasing. And releases are very important," \\
• "The thing is that I want to see that this is updated regularly." \\
• "The first thing I do is go to GitHub and check the amount of contributors and how active it is maintained" \\
• "I don't like it either when nothing has changed over the past 2 years, or only 1 file at the time" \\
\begin{tabular}[c]{@{}l@{}}• "How many commits there are, when the last commit was and how active it still is. also take a look at the git issues \\ and how they are handled"\end{tabular} \\ \hline
\textbf{Amount of contributors} \\ \hline
• "Also the amount of contributors is important" \\
• "The first thing I do is go to GitHub and check the amount of contributors and how active it is maintained" \\
• "If there are many users, but only 2 contributors I do not like it. Since that comes with big risks" \\
• "If there are many contributors, that are also bonus points" \\ \hline
\textbf{Contributor process} \\ \hline
\begin{tabular}[c]{@{}l@{}}• "Another aspect is to know what kind of people are involved in the project, have they done project like this \\ in the past or is it their first time, and how easy is it to become involved"\end{tabular} \\
\begin{tabular}[c]{@{}l@{}}• "... another important thing is to try to get an idea of how easy someone can become a maintainer, and \\ have there been many over the years or is there a steady team"\end{tabular} \\ \hline
\textbf{Git Issues} \\ \hline

• "If it is on GitHub, it is certainly important how many issues are open and how often they are responded to" \\
• "The 3rd step is to look at the git issues, how many are open and what types of issues are there" \\
\begin{tabular}[c]{@{}l@{}}• "A good way to get a feeling about how serious a project is, is to look at the issue list in GitHub. There \\ you can see what the responses are and how they are being handled"\end{tabular} \\
{\color[HTML]{262626} • "And the community as well. What are the known issues, and how fast are they resolved?"} \\
• "If there are 10k issues, which do not get a response then this is a no go for me" \\
\begin{tabular}[c]{@{}l@{}}• "How many commits there are, when the last commit was and how active it still is. also take a look at \\ the git issues and how they are handled"\end{tabular} \\ \hline
\textbf{Amount of Users} \\ \hline
\begin{tabular}[c]{@{}l@{}}• "The amount of users is also a good indication, because if many people use this, it is implicitly tested. \\ Even if this was not the case before it launched. It certainly tested now in practice.\end{tabular} \\
• "If many others use it as well, that helps a lot" \\
• "An important matter is how many other people use this" \\
\begin{tabular}[c]{@{}l@{}}• "Amount of users! If I see 2 frameworks, 1 with 20 and 1 with 80 people. The choice will go to which \\ is most used.""\end{tabular} \\
• "The amount of users for example, that is important" \\ \hline
\textbf{Backing company} \\ \hline

\begin{tabular}[c]{@{}l@{}}• "Big frameworks like angular, react or vue, angular is backed by google and react by Facebook. So they\\  will go down if the company goes down and that is not going to happen anytime soon"\end{tabular} \\
• "If there is a backing company that delivers support it really gives bonus points" \\

• "For me it is really important to see the maturity of the organisation" \\
• "If a company has a certain way of developing , that is important" \\ \hline
\textbf{Ease of integration} \\ \hline
• “It is very dependant on how easy it is to implement.” \\
• "It should be easy to implement..." \\
• "If it is a small library, I look at the implementation to see if it is easy to integrate" \\
• "How easy is it to implement, what is the available documentation for that?" \\
• "a trustworthy package is one where I can install it and see if it is working within a couple of minutes. " \\
• "We take a look at how easy it is to integrate in our system, and whether or not it is secure" \\ \hline
\end{tabular}
\end{minipage}}
\label{orgadoptionfactors}
\end{table}   

\begin{table}[ht!]
\resizebox{0.81\textwidth}{!}{\begin{minipage}{\textwidth}
\hskip-1.1cm\begin{tabular}{|l|}
\hline
\textbf{Community}                                                                                               \\ \hline
• "Another thing for me is the community."                                                                       \\
• "And the community as well. What are the known issues, and how fast are they resolved?"                      \\
• "... and after that you look at things like documentation and community"                                       \\
• "After that I start to take a look at the community, to see if I know some people. \\I might have met them at a conference maybe.          \\
• "Next to reputation and community there are not a lot of things I look at, they are the 2 most important ones" \\ \hline
\textbf{Known vulnerabilities}                                                                                   \\ \hline
• "I do not even consider the package if there are a lot of cve's known for it"                                  \\
• "You can look at the history of the security vulnerabilities to get a feeling of how serious the project is"   \\ \hline
\textbf{Reputation}                                                                                              \\ \hline
• "Are other people talking about this, if it is popular that helps"                                             \\
• "Either the project or the developers can have a reputation that influences the process"                     \\
• "Next to reputation and community there are not a lot of things I look at, they are the 2 most important ones" \\ \hline
\textbf{Support}                                                                                                 \\ \hline
• "as a user of open source software you need to know how serious you are treated, and what kind of\\ support is available to you"           \\
• "If there is a backing company that delivers support it really gives bonus points"                             \\
• "The frameworks I use, I always want them to have long term support"                                           \\ \hline
\textbf{Git stars}                                                                                               \\ \hline
• "Oh and I forgot to mention GitHub stars, they are essential!"                                                 \\
\hline
\textbf{Knowledge of team}                                                                                       \\ \hline
• "If there already are people with knowledge of the package in my team,  that helps a to decide whether\\ or not it will fit in our project" \\ \hline
\end{tabular}
\end{minipage}}
\end{table}

\subsection{Economical adoption factors}
\begin{table}[ht!]
\resizebox{0.85\textwidth}{!}{\begin{minipage}{\textwidth}
\hskip-1.0cm\begin{tabular}{|l|}
\hline
\textbf{License}                                                                                        \\ \hline
\begin{tabular}[c]{@{}l@{}}• "A license can actually be very dangerous for us, there are licenses that forbid military usage, but \\ also others that require that our code needs to be open source too"\end{tabular} \\
\begin{tabular}[c]{@{}l@{}}• "We had to take a look at licenses, since if we used certain software with limited licenses, we could\\  not sell the end product"\end{tabular} \\
• "License is definitely a big factor, you do need to have a clear image on costs or limitations there" \\ \hline
\textbf{Total costs}                                                                                    \\ \hline
\begin{tabular}[c]{@{}l@{}}• "For me the economical factors play a role in deciding what package, however not in trusting the \\ package"\end{tabular} \\
• "We take a look at the pricing, open source is better since its free"                                 \\ \hline
\end{tabular}
\end{minipage}}
\label{ecoadoptionfactors}
\end{table}

\newpage
\section{Trust factors}

\subsection{Technical trust factors}
\begin{table}[ht!]
\resizebox{0.75\textwidth}{!}{\begin{minipage}{\textwidth}
\hskip-1.1cm\begin{tabular}{|l|}
\hline
\textbf{Documentation} \\ \hline
• "Good documentation definitely contributes to trusting a package, it shows me I can use the software for what I want\\  to use it for" \\
• "A good read me is crucial, has someone put real effort in that. In other words how is the documentation" \\
• "By looking at the documentation we can see if it is a good or bad package" \\ \hline
\textbf{Code quality} \\ \hline
• "especially to validate that it does what it should do tests are very important" \\
• "Not just the code quality, but also if there is a general test suite and how much it covers the whole package" \\ \hline
\textbf{Source code}                                                                                          \\ \hline
• "The good thing about open source is that the source code is often available, so I can often check the code itself.\\  This really helps with trusting the package \\ 
• "Another aspect is the quality of the code itself, sometimes you cannot read this but you can get a general feeling on the\\  code quality" \\
• "There are also several code quality tools that grade a package, they are also good indicators to watch" \\\hline

\end{tabular}
\end{minipage}}
\label{techtrustfactors}
\end{table}

\subsection{Organizational trust factors}
\begin{table}[hb!]
\resizebox{0.75\textwidth}{!}{\begin{minipage}{\textwidth}
\hskip-1.1cm\begin{tabular}{|l|}
\hline
\textbf{Amount of users} \\ \hline
• "A package that is used by 300.000 others, I trust more than a package that is released last week and has been used by \\ 5 people, my trust in that is way lower even though it might be better that the bigger one" \\
• "If many others use it as well, that helps a lot" \\
• "Amount of users! If I see 2 frameworks, 1 with 20 and 1 with 80 people. The choice will go to which is most used."" \\
• "The amount of users for example, that is important" \\
• "A good indication on how active the project is used is the weekly downloads for example" \\
• "After that I look at the size of the community and other users... If there are enough other users, it means there are \\ enough people who can also experience problems that need fixing" \\ \hline
\textbf{Community} \\ \hline
• "I think community is very important, it should be actively used" \\
• "I would like to add something to the previous question,  namely an active mailing list around a library actually does \\ gives me more trust  in the library or package" \\
• "Another thing for me is the community." \\
• "And the community as well. What are the known issues, and how fast are they resolved?" \\
• "Things that really give me a feeling of trust are, are there smart developers involved, how are the discussions online\\  and to the developers give reasoning why they do certain things" \\
• "How large the community is says a lot" \\
• "After that I look at the size of the community and other users..." \\ \hline
\textbf{Contributors} \\ \hline
• "... and if a project is maintained by 1 or two people, the chance that the project stops due to issues in their personal life \\ are quite real. So for me, community really gives trust if it is a large and stable community." \\
• "If a project with 3000 contributors, however 99,9 of the code is written by 1 person than the amount of contributors \\ still doesn't tell me anything, so I will have less trust for that project" \\
• "Things that really give me a feeling of trust are, are there smart developers involved, how are the discussions online \\ and to the developers give reasoning why they do certain things" \\
• "Try to get an idea of how hard it is to become a contributor, over the last years several back doors have appeared in \\ libraries because it was to easy to become a contributor for example" \\ \hline

 \hline
\end{tabular}
\end{minipage}}
\label{orgtrustfactors}
\end{table}
\newpage

\begin{table}[ht!]
\resizebox{0.78\textwidth}{!}{\begin{minipage}{\textwidth}
\hskip-1.4cm\begin{tabular}{|l|}
\hline
\textbf{Active maintenance}                                                                                   \\ \hline
• "My trust in a package is mainly from GitHub, so the stars, commit history and the open and closed issues"  \\
• "They are constantly updating and releasing. And releases are very important,"                              \\
• "The thing is that I want to see that this is updated regularly."                                           \\ \hline
\textbf{Reputation}                                                                                           \\ \hline
\begin{tabular}[c]{@{}l@{}}• "I work a lot with PHP, and there are some well known companies and people in the community. If you see a\\  project is guided or written by one of them you know you are in the clear. That ensures more trust instantly"\end{tabular} \\ \hline \\
• "If you can read what the code doe exactly that is important"                                              \\
• "I think that being able to see the source code really creates trust"                                       \\
• "Like I previously said, it first has to do what I need it to do before I look any further"                 \\ \hline
\textbf{Ease of integration}                                                                                  \\ \hline
• "a trustworthy package is one where I can install it and see if it is working within a couple of minutes. " \\ \hline
\textbf{Stack overflow activity}                                                                               \\ \hline
• "Stack overflow activity also adds trust"                                                                    \\ \hline
\textbf{Git Issues} \\ \hline
• "My trust in a package is mainly from GitHub, so the stars, commit history and the open and closed issues" \\
• "I have seen several teams that used a project and then stopped using it after a while because the issues were not resolved,\\  so I think that is really important for trusting a project" \\
• "I look at everything from Git Issues, til meetups and how the community communicates" \\ \hline
\textbf{Git Stars} \\ \hline
• "My trust in a package is mainly from GitHub, so the stars, commit history and the open and closed issues" \\
• "One thing I then look at is the stars on GitHub" \\ \hline
\textbf{Backing company} \\ \hline
• "I also look if there is a commercial interest, with perhaps a company that supports this project. Red-hat is a nice \\ example of that" \\
• "If a project is backed by a good company, that also creates trust" \\
• "For me it is really important to see the maturity of the organisation" \\
• "If a company has a certain way of developing , that is important" \\ \hline
\end{tabular}
\end{minipage}}
\end{table}

\section{Trust metrics}
\subsection{Technical trust metrics}

\begin{table}[ht!]
\resizebox{0.82\textwidth}{!}{\begin{minipage}{\textwidth}
\hskip-1.4cm\begin{tabular}{|l|}
\hline
\textbf{Complexity}                                                                                                                  \\ \hline
• "A metric to determine this is the cyclomatic complexity, or how many lines of code in general. I think that would\\ be a good start." \\
• "Another aspect would be to get an indication how complex it is, and how many people use the full complexity"  
                   \\\hline    
\textbf{Tests} \\                                                                                                 \hline                     
• "Would a metric like test coverage be a solid one?"                                                                                  \\
• "Is there something that can prove that it is functional, and how is the test coverage in that?"                                     \\
• "What is important is basically, you can see with many package managers I think these build scripts that are run to\\ test this. And to validate this. \\ 
• "Yes you can see how many times has it passed, and how many times has it failed. And I think these are important\\ metrics to calculate trust. \\ \hline
\end{tabular}
\label{techtrustmetrics}
\end{minipage}}
\end{table}
\newpage

\begin{table}[ht!]
\resizebox{0.8\textwidth}{!}{\begin{minipage}{\textwidth}
\hskip-1.2cm\begin{tabular}{|l|}
\hline
\textbf{Code quality}                                                                                                \\ \hline
• "I think it is hard because it is about trust, however I think code quality is the most important in this one"                       \\
• "If we then have the possibility I would like to see some code quality metrics, dependant on the ecosystem."                         \\
• "This does fall under code quality however does deserves to be named separately is the documentation."                               \\ 
• "We would then be discussing activity and security metrics... For security and quality I would grasp to some\\ models I already know from the Software Improvement group."
\\ \hline
\end{tabular}
\end{minipage}}
\end{table}

\subsection{Organizational trust metrics}

\begin{table}[hb!]
\resizebox{0.76\textwidth}{!}{\begin{minipage}{\textwidth}
\hskip-1.4cm\begin{tabular}{|l|}

\hline
\textbf{Active maintenance} \\ \hline
• "It is not about the bugs, but how fast they are fixed. This way we can see that there are active patches coming out." \\
So how often is it updated? This is a tricky one since if a certain package is released and it is a very basic function\\  but it works, then why would you update it? \\
• "So how would you then define the number of releases. It is a hard metric." \\
• "Yeah so some measurable aspects then, this could be the last release and amount of commits per year I think." \\
• "In addition, the last release date and perhaps some data on how often there are new releases" \\
• "In addition I would also like to see it actively maintained, unless it is such a fundamental package that is does not \\ need maintenance." \\ \hline
\textbf{Amount of contributors} \\ \hline
 • "I think users, contributors, releases. " \\
• "I think community size, with that I mean 2 aspects: the amount of contributors and the amount of users that use \\ that component" \\
• "I would like to see the amount of stars, the amount of contributors and amount of weekly installs." \\ \hline
\textbf{Users} \\ \hline
• "I think community size, with that I mean 2 aspects: the amount of contributors and the amount of users that use \\ that component"\\
• "Who made it and how many people are using it?" \\
• "A really cool thing would be the ratio between people who try to use it and 3 months later still do" \\
• "I would like to see how many people use the component and for how long they have used it" \\
• "I think users, contributors, releases.  " \\
• "I would like to see the amount of stars, the amount of contributors and amount of weekly installs." \\
• "I would like to see that it is active, more specific actively used" \\ \hline
\textbf{Stability metrics} \\ \hline
• "So what would be nice to see is something to cover the stability, so how many incidents have happened with\\  a certain version." \\
• "If we look at open source projects, I would like to see something on the stability of the team" \\
• "Another aspect is how diverse their revenue streams are, if they are relying on 1 source this could fall apart more\\  easily" \\
• "You should mention the supported frameworks, platforms and operating systems" \\ \hline
\textbf{Git issues} \\ \hline
• "Most of the times there is no response to the issues, so if those are responded to that already tells a great deal" \\
• "If we could get a score on how fast issues on GitHub are closed, that would score points" \\
• "I think that the amount of GitHub stars as well as the amount of issues that are open" \\
• "I think the ratio between the amount of closed and open issues would be a nice addition" \\
• "I would like to see the amount of stars, the amount of contributors and amount of weekly installs." \\
• "I think they also did some measurements for an issue resolution time, that would be a great addition" \\ \hline
 \\ \hline
\end{tabular}
\label{orgtrustmetrics}
\end{minipage}}
\end{table}

\begin{table}[ht!]
\resizebox{0.8\textwidth}{!}{\begin{minipage}{\textwidth}
\hskip-1.4cm\begin{tabular}{|l|}
\hline
\textbf{Known vulnerabilities} \\ \hline
• "At first we want to know that there are few known vulnerabilities" \\
• "Another thing to note is the amount of CVE's that are known" \\
• "Some statistics and history on prior security bugs" \\
• "Look if there are known vulnerabilities I would like to know, even if it's that simple, it's still something already" \\ \hline
\textbf{Stackoverflow presence} \\ \hline
• "If we may fantasize about it a little I think that the amount of questions on stackoverflow would be a great \\ addition, as well as blog posts or presence on reddit." \\ \hline
\textbf{Reputation} \\ \hline
• "What important is for me, is that e.g. someone like Linus Torvalds expresses faith in the project \\
• "Reputation of the developer is a good metric as well" \\ \hline
\textbf{Developing entity} \\ \hline
• "I would like to see who is behind it, the person or company. Even though it then is difficult to find out if you \\ can trust it but you can at least try." \\
• "It can very well be that you do not get any information out of it but let's take react for example, that is created\\  by Facebook. Even though I do not like Facebook as a company, I do trust that the software they produce is good. \\
• "Reputation of the developer is a good metric as well"  \\  \hline

\end{tabular}
\end{minipage}}
\end{table}

\newpage
\section{Odyssey Momentum}

The massive online collaboration event or hackathon as it is normally called was a three day event focused on creating projects and sharing knowledge. There were 13 different tracks that each covered different social or technical problems. A team could then reach out to the organisation with a possible solution, and this weekend then provided the opportunity to start creating. This weekend had a very large focus on the collaboration aspect. This is a yearly physical event and due to the corona virus could not be held this year. Instead the creators created this online collaboration platform in which one could roam freely and socialize with other participants of the event. This way if a team was stuck on a certain problem, they could fly around in the Odyssey-world and find some other participant who does have the knowledge to solve this issue. This is a beautifully designed concept that allowed teams to collaborate remotely whilst working on their own projects. Since my coding skills are not nearly as good as my team members'. I was part of the 'marketing team' for this weekend. This meant keeping the social media platforms up-to-date with the team's progress. It also meant communicating with other teams to find the appropriate other participants that could help us solve issues that we had. There also were several presentations that were done by the marketing team. Overall this was an amazing addition to the research project. Even though I did not code too much of the actual prototype, this weekend was a great learning experience in terms of communicating within a team and being the middle man for communication towards our team and the organization. The cherry on top is that we won the track in which we competed! This was an amazing finalization of the weekend!\\

\bibliographystyle{apalike}
\bibliography{Chapters/bib.bib}

\end{document}